\documentclass[%
article,
 floatfix,
 aps,
 prb,
 groupedaddress,
preprint,
 amsmath,amssymb,
floatfix,
]{revtex4-2}

\usepackage{graphicx}
\usepackage{dcolumn}
\usepackage{bm}
\usepackage{upgreek}
\def\Mu{\upmu}

\begin{document}

\title{Stabilization and application of asymmetric N\'eel skyrmions in hybrid nanostructures}

\author{Mateusz Zelent}
\affiliation{Institute of Spintronics and Quantum Information, Faculty of Physics, Adam Mickiewicz University, Poznan, ul. Uniwersytetu Poznańskiego 2, Poznan, PL-61-614 Poland}

\author{Mathieu Moalic}
\affiliation{Institute of Spintronics and Quantum Information, Faculty of Physics, Adam Mickiewicz University, Poznan, ul. Uniwersytetu Poznańskiego 2, Poznan, PL-61-614 Poland}

\author{Michal Mruczkiewicz}
\affiliation{Institute of Electrical Engineering, Slovak Academy of Sciences, Dubravska cesta 9, SK-841-04 Bratislava, Slovakia}
\affiliation{Centre For Advanced Materials Application CEMEA, Slovak Academy of Sciences, Dubravska cesta 9, 845 11 Bratislava, Slovakia}

\author{Xiaoguang Li}
\affiliation{
College of Engineering Physics, Shenzhen Technology University, Shenzhen 518118, China}

\author{Yan Zhou}
\affiliation
{School of Science and Engineering, The Chinese University of Hong Kong, Shenzhen 518172, China}

\author{Maciej Krawczyk}
\affiliation{Institute of Spintronics and Quantum Information, Faculty of Physics, Adam Mickiewicz University, Poznan, ul. Uniwersytetu Poznańskiego 2, Poznan, PL-61-614 Poland}

\date{\today}

\begin{abstract}
Increasing amounts of information force the continuous improvement of information storage and processing technologies, further device miniaturization, and their efficiency increase. Magnetic skyrmions, topological quasiparticles, and the smallest stable magnetic textures possess intriguing properties and potential for data storage applications. Hybrid nanostructures with elements of different magnetization orientations can offer additional advantages for developing skyrmion-based spintronic and magnonic devices. We show that an  N\'eel-type  skyrmion confined within a nanodot placed on top of a ferromagnetic stripe produces a unique and compelling platform for exploring mutual coupling between magnetization textures. The skyrmion induces an imprint upon the stripe, which, in turn, asymmetrically squeezes the skyrmion in the dot, increasing their size and the range of skyrmion stability for small values of DMI, as well as introducing skyrmion bi-stability. At the end, we present a proof-of-concept technique for unconstrained transport of a skyrmion along a racetrack based on proposed hybrid systems. Our results demonstrate a hybrid structure that is promising for applications in magnonics and spintronics.
\end{abstract}

\maketitle
\section*{Introduction}
Magnetic skyrmions are stable nanometric-size spin textures with potential for memory, spintronic and magnonic applications due to the unique properties governed by their nontrivial topology~\cite{Moreau-Luchaire2016AdditiveTemperature,Marrows2015,Kiselev_chilar_skyrmions,Zhang2020Skyrmion-electronics:Applications,Zhou2019MagneticConcepts,Finocchio2021MagneticApplications}. The chiral term of energy, known as Dzyaloshinskii-Moriya interactions (DMI), which originates from spin-orbit coupling and inversion symmetry breaking in single-crystal materials like B20-type crystals (bulk DMI) or in thin ferromagnetic films at the interface with heavy metal (interfacial DMI), are essential for the stabilization of the Bloch and N\'eel type skyrmions, respectively, additionally to the perpendicular magnetic anisotropy (PMA) in the case of thin films~\cite{Wang2018ASize,Yang2015AnatomyInterfaces, Behera2018SizeAnisotropy}. Both the static and dynamics properties of skyrmions are intensively studied in a wide range of materials and structures~\cite{Kiselev_chilar_skyrmions,Zhang2016b,Zhang2020Skyrmion-electronics:Applications,5}. In particular, the stability of skyrmions at room temperature has been demonstrated in thin films and in confined geometries, like ferromagnetic stripes and nanodots~\cite{Yu2016Room-TemperatureAsymmetry, Tejo2017, Saha2019FormationAnisotropy, PSSR:PSSR201700259}. Very often, in spite of the topological protection, the skyrmion configuration is not self-nucleating and favorable energetically ground state~\cite{Vetrova2021InvestigationDot,Riveros2021Field-DependentNanodots,Rajib2021RobustDMI}.
 Therefore, an important question arises whether the right selection of materials, designing their structure, or changing environmental conditions can improve the stability of skyrmion, help to obtain effective control over it, which would increase also its functionality.
 
 The synthetic antiferromagnets offer an important advance in skyrmions' principal application, i.e., as an information carrier in race-track memory. In such bilayered tracks with opposite magnetization polarization, the pair of coupled skyrmions exerts opposite torques, allowing to mitigate skyrmion spin-Hall effect (SHE) and providing a straight flow of the skyrmions\cite{Zhang2016}. From that application point of view, also elliptically deformed skyrmions show promising properties~\cite{Hsu2016,Jena2020EvolutionSymmetry,Xia2020}. Here, the axis-symmetry breaking has been achieved by an in-plane bias magnetic field~\cite{Zhang2020DeformationMicroscopy} or by introducing a certain in-plane anisotropy~\cite{Shibata2015LargeCrystal}, for instance, by applying an in-plane strain~\cite{Shibata2015LargeCrystal,Hagemeister2016}. Recent studies show that also an anisotropic PMA and DMI may stabilize elliptical N\'eel-skyrmions in thin films~\cite{Cui2021,Camosi2021}. 

\begin{figure}[!ht]
 \includegraphics[width=1.0\textwidth]{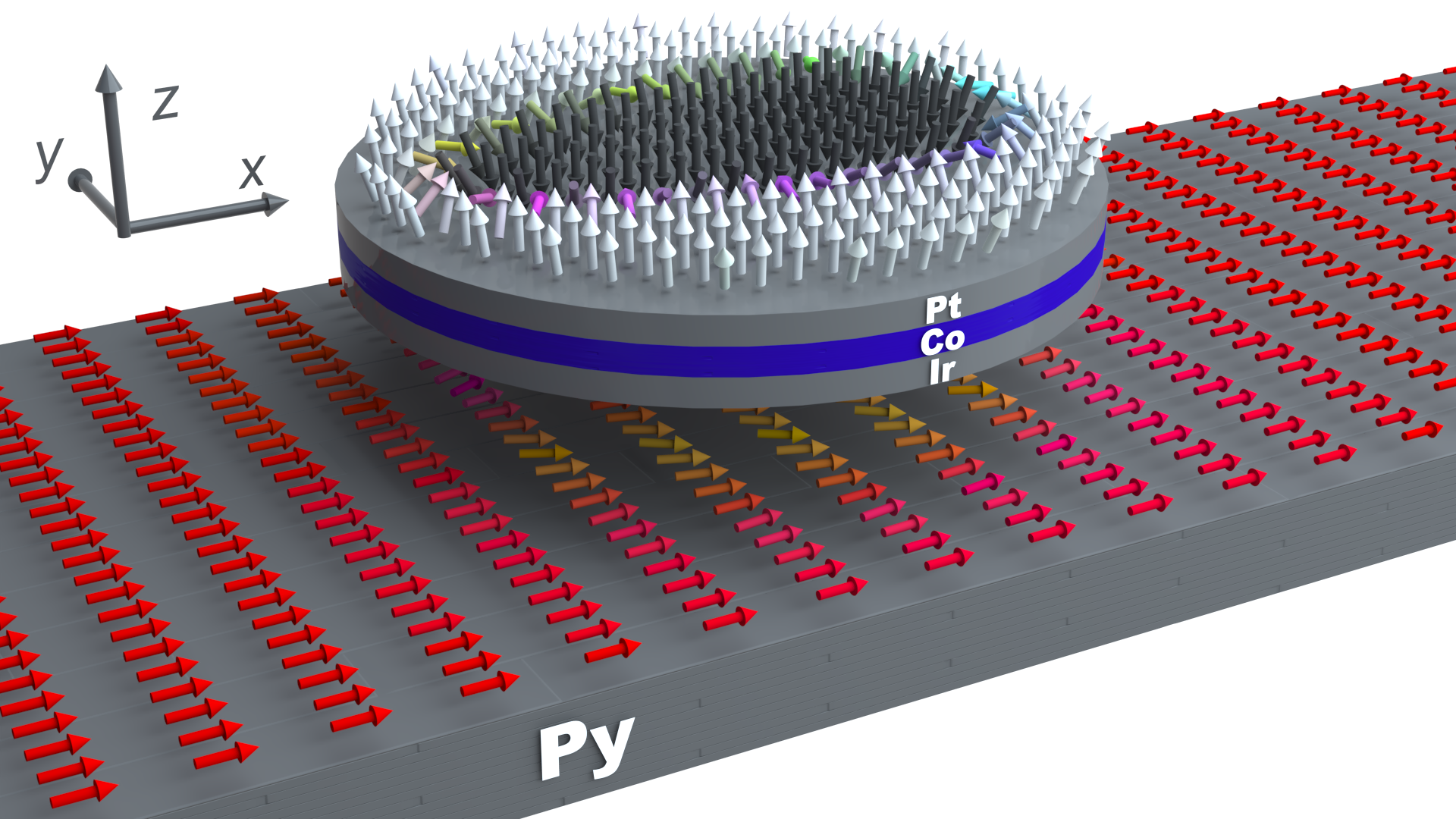}
 \caption{Visual representation of the system under consideration. The Pt/Co/Ir multilayer dot is located slightly above the Py stripe. In the dot an egg-shape Néel-type skyrmion state is stabilized by the magnetostatic coupling to the skyrmion imprint upon the in-plane magnetized stripe. The arrows indicate the direction of magnetization. 
 }
 \label{fig:static_anim}
\end{figure}

 We show that deformed, an oval shape with one axis of symmetry (egg-shaped like), N\'eel skyrmions can be stabilized by magnetostatic interaction in a hybrid structure composed of a multilayered nanodot hosting a skyrmion and the in-plane magnetized thin stripe made of soft ferromagnetic material, Fig.~\ref{fig:static_anim}. The skyrmion state generates a nonuniform stray magnetic field, which affects the magnetization in the adjacent layer. The disturbed magnetization order in the adjacent layer induces a reverse counter--stray magnetostatic field, which exerts a significant effect on the skyrmion static configuration, breaking its circular symmetry and enhancing an influence of DMI. Importantly, this mutual interaction of the skyrmion and the stripe increases the skyrmion stability and opens a narrow range of the DMI parameter for bi-stabilization of the skyrmion. Moreover, the interaction results also in an increase in skyrmion size, analogous to the effect of an external magnetic field~\cite{Wang2018ASize,Tejo2017,Aranda2018MagneticInteraction,Srivastava2020ObservationPtMnGa}, thus allowing for skyrmion size control. Furthermore, the strength of the skyrmion's asymmetric deformation can be controlled by DMI strength, magnetic anisotropy of the film, the skyrmion or stripe polarity, and the skyrmion chirality. There is an important role of the nanodot edge in the skyrmion stabilization, but we found that an elongation of the nanodot along the direction perpendicular to the ferromagnetic stripe magnetization preserves an egg-shape skyrmion, providing a suitable geometry for testing skyrmion flow, and opening potential for utilization in race-track memories. Thus, this research opens an avenue for controlling the properties of topological objects in hybrid structures for spintronic and magnonic applications.
 
The present paper is organized as follows. First, we analyze the skyrmion size and shape in dependence on the DMI value, then we analyze the magnetostatic fields induced by skyrmions in the isolated nanodot and the coupled system. Next, we discuss how the imprint in the stripe affects the shape of the skyrmion in the nanodot. Finally, we present a numerical proof-of-concept of the skyrmion moving technique, which uses the specific properties of the egg-shaped skyrmion in a hybrid system to reduce the impact of the skyrmion Hall effect on a racetrack.
 
\section*{Results and discussion}
\subsection*{Results} 

We investigate a circular nanodot with the thickness of 1.5 nm and 150 nm radius, positioned directly 1.5 nm above a ferromagnetic stripe (see Figure~\ref{fig:static_anim}) magnetized along the positive $x$-axis. The stripe is 384 nm wide, 4.5 nm thick and 6 $\upmu$ m long. The magnetic dot is defined with an effective thickness approach as a Pt/Co/Ir multilayer structure~\cite{PSSR:PSSR201700259} with DMI and PMA, which form favorable conditions for the stabilization of a skyrmion. For the stripe material, we assume Py. The magnetic parameters of the multilayer and Py are taken from the literature~\cite{PSSR:PSSR201700259,Moreau-Luchaire2016AdditiveTemperature} and are listed in Method section.
 
We perform micromagnetic simulations using Mumax3~\cite{MuMax2011_main,mumax_2014,Leliaert2018FastMumax3} to investigate the steady state in the hybrid system, but first, we extract the DMI range of the skyrmion stability in an isolated Pt/Co/Ir nanodot. We found a stable circular skyrmion state for the absolute DMI (|DMI|) value from 0.86 mJ/m$^2$ to 1.75 mJ/m$\mathrm{^2}$, with skyrmion diameter increasing monotonously from 7 nm to 215 nm, [see, the blue dashed line in Figure~\ref{fig:d_dependence} (a,b)], regardless of the DMI sign. In the range of |DMI| from 0.86 mJ/m$^2$ to 1.1 mJ/m$^2$ we observe a very small skyrmion with a slight increase of its diameter from 7 to 35 nm with increasing DMI. At |DMI| = 1.1 mJ/m$^2$, the rapid increase of the diameter is observed and a large skyrmion stabilize, reaching the value of 215 nm at 1.75 mJ/m$^2$. 

\begin{figure}
    \includegraphics[width=0.7\textwidth]{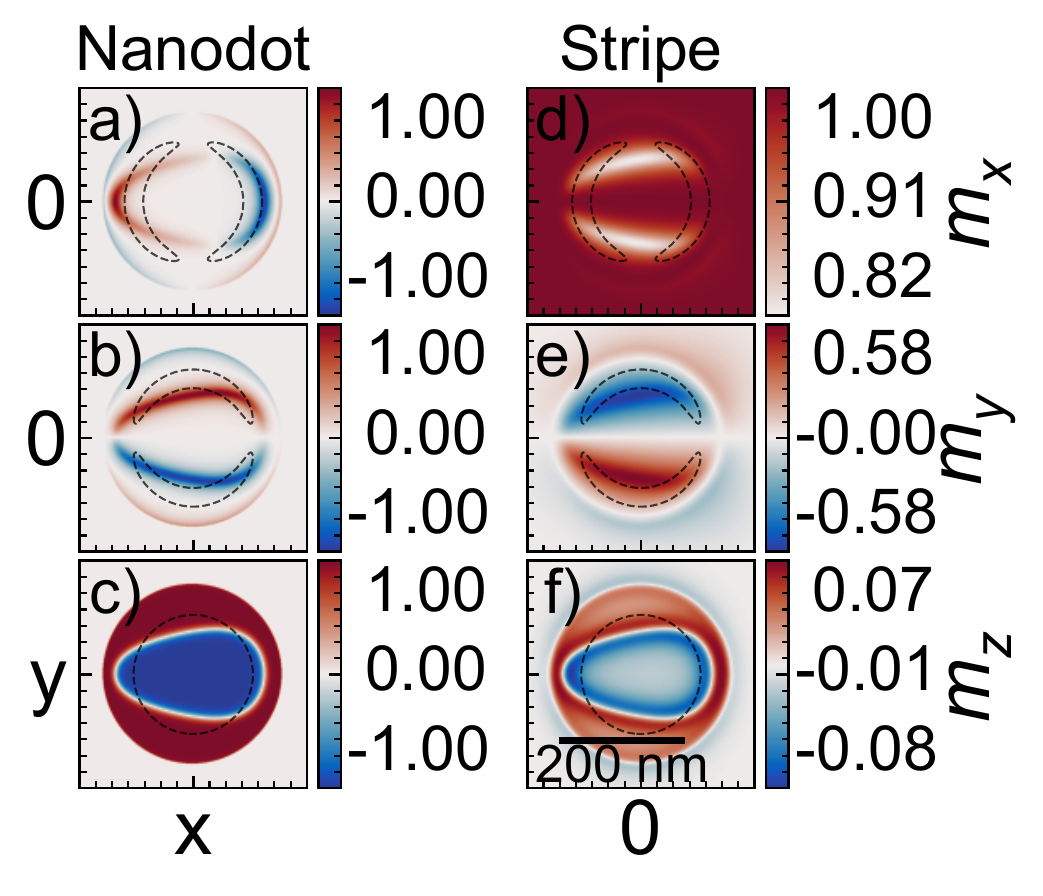}
    \caption{Static magnetization configuration in a Pt/Co/Ir nanodot with DMI = 1.6 mJ/m$^2$ dipolarly coupled to the Py stripe. The color scale is given in reduced units of magnetization. (a-c) The $m_x$, $m_y$, and $m_z$ components of magnetization in the middle plane of the nanodot, and (d-f) the components of the magnetization in the stripe. The dashed line was plotted as an isoline of 75\% of the maximum positive or negative magnetization amplitude (a,b,d,e) and $m_{z}=0$ (c,f) of the skyrmion texture for the isolated system, as an approximation of the shape and size of the reference system. 
    }
    \label{fig:single_static}
\end{figure}

In the next step, we perform relaxation simulations of a full system, a nanodot coupled with a stripe. The shape anisotropy of the stripe maintains the uniform magnetization along the $x$-axis, while the spacer between the elements guarantees coupling only by the magnetostatic stray field. For DMI equal to $-1.6$ mJ/m$\mathrm{^2}$ we observed after relaxation strongly deformed, like an egg-shape, skyrmion (see Fig.~\ref{fig:single_static} (a-c)) with the maximum size $s_{y}= 150$ nm and $s_{x}= 240$ nm, along the ${y}$ and $x$-axis, respectively. Skyrmion size is measured as a distance between the vertically or horizontally furthest points of the skyrmion domain wall, where $m_{z}=0$ [see Fig.~4(e)]. 

In Fig.~\ref{fig:single_static}, the black dashed lines show the contours of the skyrmion domain wall in the isolated nanodot (200 nm in diameter). Note that the observed egg-shape of the skyrmion has the size along the $y$-axis near the left edge of the nanodot ($x<0$) smaller then near the right edge ($x>0$), which is correlated to the magnetization orientation in the stripe, as we will show in the following part. There is also significant difference in all magnetization components with respect to the skyrmion in the isolated nanodot.

\begin{figure}
\includegraphics[width=1.0\textwidth]{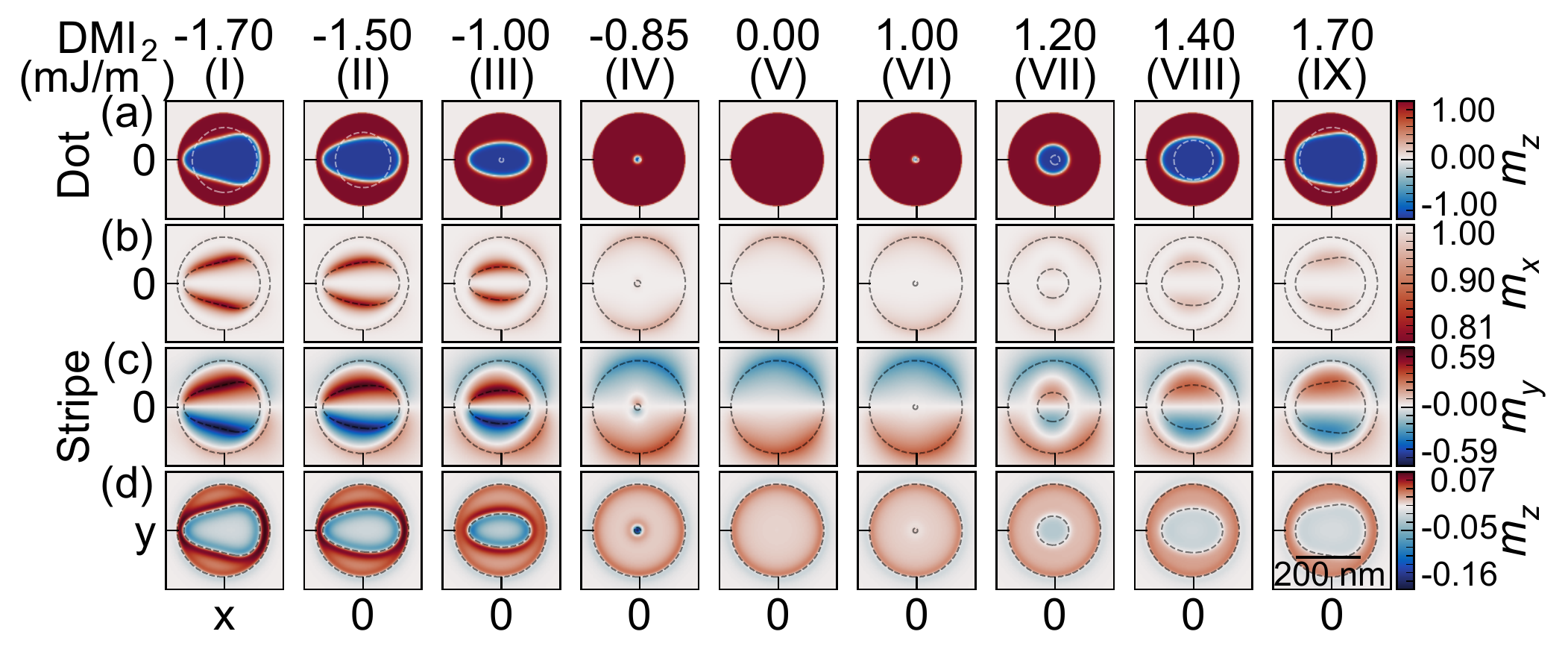}
 \caption{Images of the relaxed magnetization configuration of the Pt/Co/Ir nanodot over the Py stripe for different values of the DMI in a nanodot. (a) The out-of-plane component of the magnetization inside the nanodot. The white-dashed line marks the shape of the skyrmion in an isolated nanodot at respective DMI.
(b-c) The in-plane and (d) the out-of-plane components of the magnetization inside the stripe--the skyrmion's imprint. The dashed lines represent the edge of the nanodot (outer circle) and the edge of the skyrmion (inner circle).
}
\label{fig:images_dmi_depenece}
\end{figure}

Figure~\ref{fig:single_static} (d-f) shows the deformation of the magnetization components in the stripe from the ${\bf m}=({m_x},0,0)$ alignment. The magnetization in the stripes is deviated from their equilibrium position in the areas of interaction with the edge of the nanodot and the magnetic texture of the skyrmion, and is further called an \textit{imprint}. A larger disturbance of the magnetization is observed on the $m_{y}$ (up to $\pm$ 0.58) then the $m_{x}$ component (up to 0.18). The reduction of the $m_{z}$ component reaches $\pm 8\%$ at the horizontal edges of the skyrmion. Moreover, all magnetization components are asymmetric with respect to the vertical midplane of the nanodot, but antisymmetric with respect to the horizontal midplane of the system. 

The skyrmion and the imprint dependence on the sign and strength of the DMI is shown in Fig.~\ref{fig:images_dmi_depenece}. We are varying the value of DMI interactions from $ \pm $ 0.86 mJ$/m^2$ (the skyrmion formation) to ~$\pm$ 1.75 mJ/m$^2$ (skyrmion instability). We show the $m_{z}$ component of the magnetization in the nanodot in (a), and $m_{y}$, $m_{x}$ and $m_{z}$ components in the stripe in (b-d). The most important observations regarding the changes in a nanodot are following: (i) The significant increase of the skyrmion size as compared to the isolated dot. 
(ii) The stronger increase in a skyrmion size and larger egg-shape deformation at negative then positive DMI values. 
The changes in the stripe magnetization are:
(iii) The values of all magnetization components in the imprint are lower for a positive then a negative DMI. The maximal intensity of the $m_{y}$ component is about $\pm $0.59 for the imprint at DMI = -1.7 mJ/m$\mathrm{^2}$, while for the positive DMI it is only $ \pm $ 0.17 at the normalized magnetization units.
(iv) For all considered DMI values, the imprint retains the same type of symmetry and asymmetry as the  skyrmion in the nanodot. 
(v) The strength of the imprint induced from the edge of the nanodot becomes weaker as the skyrmion grows. 

\begin{figure}[!htp]
\includegraphics[width=1.0\textwidth]{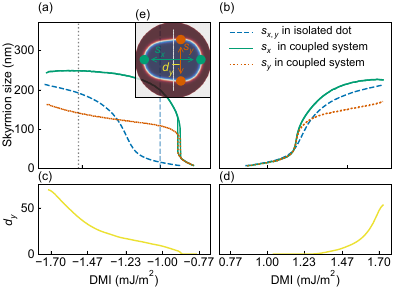}
\caption{The simulated maximal skyrmion size along the $x$ and $y$ axis in dependence on the DMI strength, (a) between -1.75 to -0.75 mJ/m$^2$ and (b) between 0.75 to 1.75 mJ/m$^2$. The blue dashed line represents diameter for an isotropic circular skyrmion in the isolated nanodot. Green $s_x$ and orange $s_y$ curves represent the skyrmion maximum size measured along the $x$ and $y$ axis, respectively, in the coupled system. The vertical, light-blue line indicates the DMI value for which the analysis of the effect of magnetocrystalline anisotropy on skyrmion size is performed in Fig.~\ref{fig:sub-units}(d). (c,d) The measure of the skyrmion asymmetry with respect to the $y$-axis, i.e., the shift of the maximum size $s_y$ of the skyrmion from the symmetry position $x=0$, $d_y$. (e) Image of the relaxed skyrmion for DMI = 1.6 mJ/m$^{2}$, with the markings of the plotted values.}
\label{fig:d_dependence}
\end{figure}

The quantitative analysis of the skyrmion change with increasing DMI is shown in Fig.~\ref{fig:d_dependence}. For the positive sign of the DMI, up to 1.1 mJ/m$\mathrm{^2}$, we observe no difference between the skyrmion size and shape of the isolated nanodot and the nanodot coupled with stripe, but above this value the skyrmion in a hybrid structure deforms and becomes larger than the skyrmion in an isolated nanodot. For negative DMI values, the described effects are significantly stronger, causing a sharp increase of the skyrmion at smaller DMI values. Importantly, the skyrmion starts to stabilize at lower DMI values compared to the isolated dot, i.e., at -0.788 mJ/m$\mathrm{^2}$ instead of 0.863 mJ/m$^\mathrm{^2}$. The sudden skyrmion growth happens also at significantly smaller DMI magnitude in the hybrid structure than in the isolated nanodot, i.e., at -0.82 mJ/m$^2$ instead of -1.23 mJ/m$^2$ in the nanodot.

Interestingly, in a narrow range of negative DMI values, from -0.82 to -0.88 mJ/m$\mathrm{^2}$, we observe bi-stability of the skyrmion~\cite{PSSR:PSSR201700259,Beg_2015_Ground_state_search_hysteretic_behaviour,Tejo2017}, i.e., a simultaneous occurrence of two possible realizations of the stable skyrmion state, distinguished by different skyrmion size, around 25 nm and 125 nm at --82 mJ/m$\mathrm{^2}$. For higher negative DMI values, we observe a significant disparity between the sizes of the skyrmion along the $x$ and the $y$ axes ($s_x$ and $s_y$), as well as a strong asymmetric deformation of the skyrmion along the $x$-axis ($d_y$--the shift of the maximum skyrmion size along the $x$-axis from the symmetrical position), see Fig.~\ref{fig:d_dependence}(a,c). This disparity reaches even 100 nm at --1.25 mJ/m$\mathrm{^2}$. For each value of the DMI presented here, the topological number of the skyrmion is preserved and is invariably around $\pm$~1. 

\newpage

\subsection*{Discussion}

Although a long stripe with saturated magnetization along its axis exerts a negligible stray magnetostatic field on the nanodot (due to the finite length of the stripe), we found a large difference in the shape and size of the skyrmion in the hybrid system compared to the isolated nanodot. It is clear that the magnetostatic stray field, $\mathrm{\bf{H}}_\textrm{s}$, is responsible for these mutual interactions between the two systems and the observations described in the previous section. It is the stray field from the nanodot with the skyrmion $\mathrm{\bf{H}}_\textrm{s-dot}$ forms an imprint upon the stripe, which in turn induces a magnetostatic stray field $\mathrm{\bf{H}}_\textrm{s-str}$ in the nanodot. To explain these mutual interactions, we will analyze both contributions separately in the following subsections. 

\subsection*{Stray magnetostatic field from the nanodot}

\begin{figure}[!htp]
\includegraphics[width=1.0\textwidth]{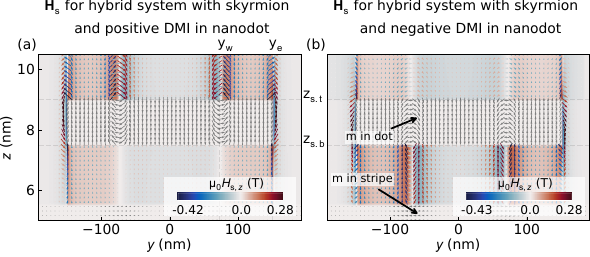}
\caption{ The spatial distribution of the stray magnetostatic field in the studied system, for the positive value of DMI, 1.6 mJ/m$\mathrm{^2}$ (a), and the negative --1.6 mJ/m$\mathrm{^2}$ (b). The plot is taken in the $(y,z)$ mid-plane cross-section of the nanodot.
 The size of the arrow is proportional to the magnetic field strength. The central and bottom part of the figure, marked by the white color and gray arrows represent the magnetization in the nanodot and the stripe. The color scale represents the intensity of the $z$-component of the $\mathbf{H}_{\mathrm{s}}$ magnetostatic field.
}
\label{fig:stray_field_3}
\end{figure}

We performed analysis of the stray magnetostatic field distribution and magnetization configurations at DMI =$ \pm $1.6 mJ/m$^2$.\footnote{For this purpose, we have performed an analysis of the magnetostatic field distribution of the system under study with reduced size of the elementary cell, just to obtain precise spatial distributions of the field. We used here 0.75 $\times$0.75 $\times$0.15 nm cell size.} Figure~\ref{fig:stray_field_3}(a) presents the total spatial distribution of the stray magnetostatic field $\mathrm{\bf{H}}_\textrm{s}$ in the coupled system for the positive DMI. Intensity of the magnetostatic field from the nanodot, $\textbf{H}_\textrm{s-dot}$, is significantly larger above ($z> z_{\mathrm{s.t}} = 9$~nm) than below ($z<z_{\mathrm{s.b}} = 7.5$ nm) the nanodot, where $z_{\mathrm{s.t}}$ and $z_{\mathrm{s.b}}$ corresponds to the top and bottom surface of the nanodot, respectively. In particular, we observe the following changes in the field strength: i) an enhancement of the fields above the skyrmion wall and the nanodot edge, ii) slightly weaker but still relatively strong field below the edge of the dot, and iii) a decrease of the stray-field value below the skyrmion wall. These properties apply to both components of the stray field, i.e., radial and vertical. Obviously, the $\textbf{H}_\textrm{s-dot}$ from the isolated nanodot has a circular symmetry.

To understand the origin of this stray field asymmetry, we involve the concept of magnetostatic surface and volume charges\cite{Aharoni2000,Coey2010}. With that we can distinguish three main contributions to $\mathrm{\bf{H}}_\textrm{s-dot}$: 
(i) $\textbf{H}_\textrm{s-dot.e}$ that originates from a tilted magnetization at the nanodot's edge, (ii) $\textbf{H}_\textrm{s-dot.dw}$, the field induced by volume charges in a N\'eel-type domain wall of the skyrmion, and (iii) $\textbf{H}_\textrm{s-dot.dm}$, the field from the surface charges of the domains, i.e., the skyrmion core and its surrounding. The field $\textbf{H}_\textrm{s-dot.dm}$ has opposite radial orientation above and below the skyrmion domain wall. The bulk charges of a given N\'eel domain wall $\textbf{H}_\textrm{s-dot.dw}$ have the same sign throughout the thickness, leading to a symmetric distribution of the stray field along the $z$-axis. Thus, the field enhancement or suppression is just a constructive or destructive superposition of these two components, $\textbf{H}_\textrm{s-dot}(x_{\mathrm{w}}, y_{\mathrm{w}}, z_\textrm{s.t/b} \pm z) = \textbf{H}_\textrm{s-dot.dw} \pm \textbf{H}_\textrm{s-dot.dm}$, where $z>0$, and ($x_{\mathrm{w}}$, $y_{\mathrm{w}}$) indicates a domain wall lateral position. Similarly, at the nanodot edge: $\textbf{H}_\textrm{s-dot}(x{_\textrm{e}},y{_\textrm{e}},z_{\textrm{s.t/b}}) = \textbf{H}_\textrm{s-dot.e} \pm \textbf{H}_{\textrm{s-dot.dm}}$ (where $x{_\textrm{e}}^2+y{_\textrm{e}}^2 = 150^2$ nm$^2$ indicates the nanodot edge position) but here $\textbf{H}_\textrm{s-dot.e} < \textbf{H}_\textrm{s-dot.dw}$, due to only slight tilt of the magnetization at the nanodot edge from the normal direction, so the effect is weaker. In fact, the described asymmetry in the stray field is a Hallbach effect, well known in permanent magnets arrangements with the rotated magnetization but here realized in a deep nanoscale with the chiral domain wall of the skyrmion~\cite{Chen2021ChiralSkyrmions,Marioni2018HalbachTextures}. Since, the sign of the DMI determines the chirality of the N\'eel domain wall, its change reverses the areas of weakening and enhancing of the stray field. Indeed, for DMI = - 1.6 mJ/m$^2$ the $\textbf{H}_\textrm{s-dot}$ field is enhanced below the nanodot as shown in Fig.~\ref{fig:stray_field_3}(b). Because $\textbf{H}_\textrm{s-dot.dw}$, $\textbf{H}_\textrm{s-dot.e} < \textbf{H}_\textrm{s-dot.dm}$,\cite{Tetienne2015TheNanomagnetometry} the orientation of the effective stray field from the nanodot, its radial and $z$ components, is always determined by the skyrmion polarization, but its strength is controlled by the domain wall chirality, i.e., the DMI sign.

 Worth to note is that the stray field induced by the skyrmion domain wall $\textbf{H}_\textrm{s-dot.dw}$ reduces the effect of the stray field $\textbf{H}_\textrm{s-dot.e}$ induced from the dot edge as a result of the opposite rotations of these fields. Thus, as the skyrmion size increases, the effective field under and above the edge of the nanodot decreases, as we have already observed in Fig.~\ref{fig:images_dmi_depenece}(b,c) (I-IX).
 Moreover, the asymmetric skyrmion deformation visible in this figure can only be related to the mutual interaction of the stripe with the stray field from the domain wall, since the edge maintains its circular symmetry. Thus, in the following sections, we will focus our discussion only on $\textbf{H}_\textrm{s-dot.dw}$ and $\textbf{H}_\textrm{s-dot.dm}$.

\newpage

\subsection*{The imprint}\label{Sec:Imprint}

\begin{figure}
 \includegraphics[width=\textwidth]{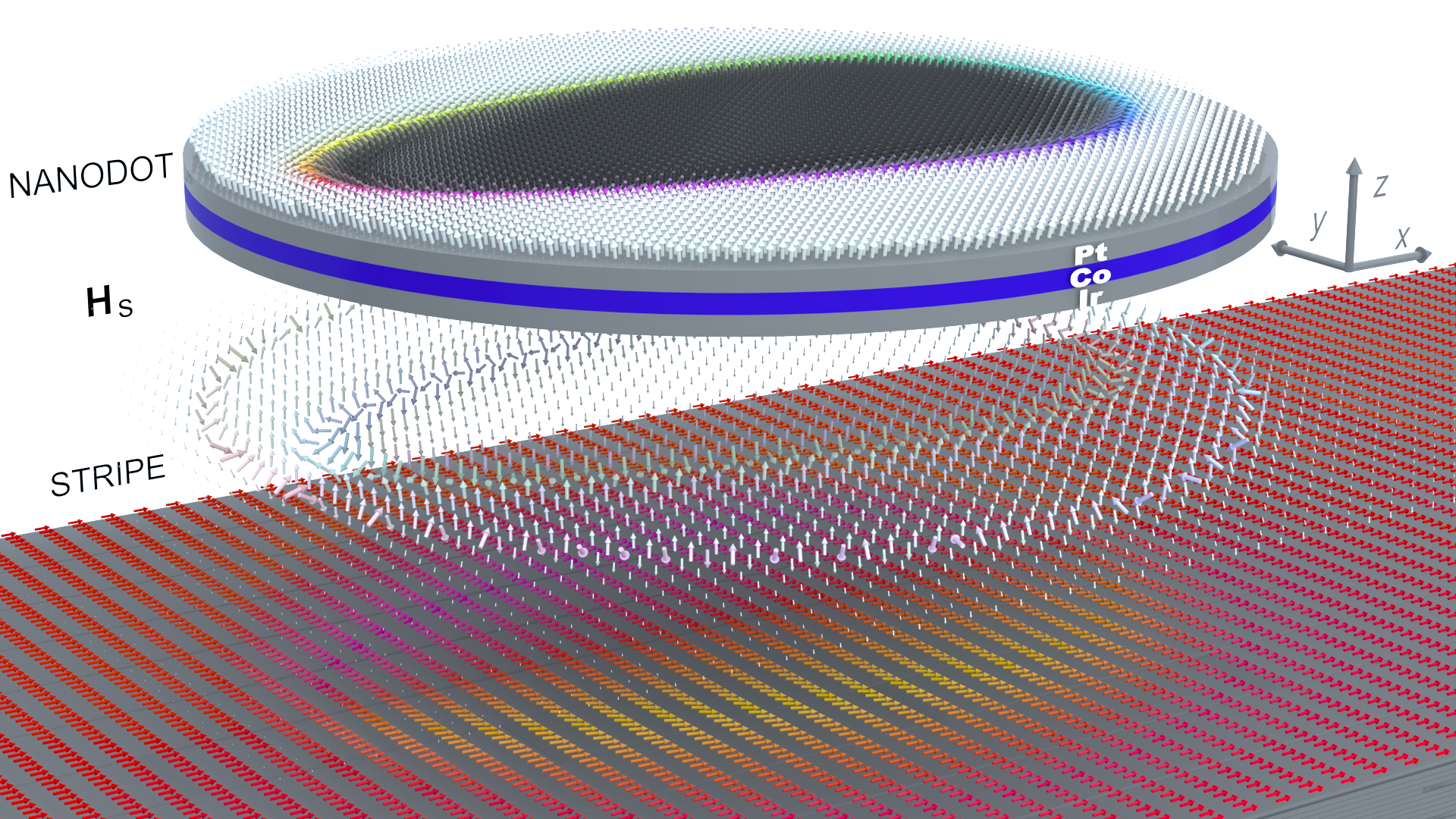}
 \caption{3D plot of the magnetization texture and the stray magnetic field distribution in the investigated system for DMI = -1.6 mJ/m$^{2}$. Figure represents spatial distribution of the magnetization in the nanodot (top), magnetic field in the space between (middle), and the magnetization in the stripe (bottom). Color of the arrows in the stripe represents the direction of magnetization.}
 \label{fig:plane_b_demag_coupled}
\end{figure}

 Figure~\ref{fig:plane_b_demag_coupled} shows the 3D plot of the magnetization configuration in the hybrid system and the field $\textbf{H}_{\textrm{s}}$ in the space between the elements for the negative value of the DMI. Below the skyrmion domain wall the effective stray field is directed centrifugal, for $x<x_{\text{w}}$ and $x>x_{\text{w}}$ gets also the $z$ component oriented according to the polarization of the nanodot domains.

These fields and their orientations, in particular $\textbf{H}_\textrm{s-dot} \approx \textbf{H}_\textrm{s-dot.dw} + \textbf{H}_\textrm{s-dot.dm}$, are important, because they exert the torque on the stripe magnetization below the skyrmion, causing its deformation, i.e., an imprint. 

\begin{figure}[!ht]
\includegraphics[width=1.0\textwidth]{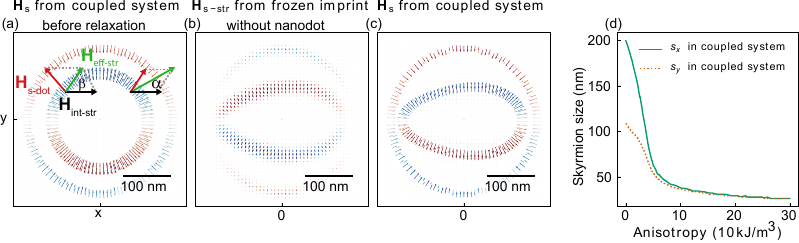}
\caption{The in-plane cross section of the magnetostatic stray-field distribution calculated in the space between the nanodot and the stripe ($z=6.5$ nm) generated by: (a) the nanodot with a circular skyrmion, (b) the imprint, and (c) in the hybrid system after relaxation. (d) The skyrmion size (its maximal dimensions along the $x$ and $y$ axis) in dependence on the magnetic anisotropy constant in the stripe. In the calculations, we assumed the DMI value --1.6 mJ/m$^{2}$ and --1.0 mJ/m$^2$ in (a-c) and (d), respectively.
}
\label{fig:sub-units}
\end{figure}

To understand the mutual interaction of skyrmion and stripe, let us first consider the demagnetizing fields of an isolated dot acting on the magnetization in an uniformly magnetized stripe. This can be considered as the first relaxation step of the coupled stripe-nanodot system, where the initial magnetic state of the stripe and nanodot is in their isolated equilibrium orientation. In that case, the initial skyrmion is fully circular, i.e., it exerts a circularly symmetric stray field. Here, the effective field in the stripe \textbf{H}$_{\textrm{eff-str}}$ at the position of domain wall can be considered as a sum of $\textbf{H}_\textrm{s-dot}({x}_{\textrm{w}},{y}_{\textrm{w}}, z < z_\textrm{s.b})$ field and the internal magnetic field of the stripe, which in our case can be limited to the shape anisotropy field:
$\textbf{H}_{\textrm{int-str}}=[\mathrm{H}_\textrm{int-str},0,0]$. During relaxation the magnetization in the stripe shall follow the direction of the effective magnetic field to minimize the torque. As schematically shown in Fig.~\ref{fig:sub-units}(a), the in-plane component of the 
effective field in the stripe $\textbf{H}_{\textrm{eff-str},\parallel}=\textbf{H}_\textrm{s-dot},\parallel+\textbf{H}_{\textrm{int-stripe}}$ is parallel to the stripe
magnetization at $y=0$ line, therefore does not exert any torque. But for $y\neq 0$ the torque is nonzero, and it has different magnitude for the halves at $x>0$ and $x<0$. Therefore, the angle between the effective field $\textbf{H}_{\textrm{eff-str},\parallel}$ and the stripe axis is larger on the left side ($\beta > \alpha$), so the magnetization tilt in the $y$-axis direction shall be stronger on the left side of the skyrmion center than on the right side. Thus, this effective field causes relaxation of the stripe magnetization to the asymmetric imprint texture. 

This analysis is confirmed in the micromagnetic simulations in Figs.~\ref{fig:single_static}(h) and \ref{fig:images_dmi_depenece}(b,c) (I-IX) for different values of DMI. It is clear that for the coupled system after the relaxation, the in-plane components of the magnetization $m_{x}$ and $m_{y}$ in the stripe are symmetric and antisymmetric with respect to the $x$-axis, respectively, and both are asymmetric with respect to the $y$-axis. 
 From our consideration, it also comes out that the change of the magnetization orientation in the stripe will reverse the skyrmion asymmetry. Moreover, the strength of the stripe's shape anisotropy shall control the strength of the magnetization imprint, and also control the asymmetry of the skyrmion, as we will show in the next section.

\newpage
\subsection*{The asymmetric deformation of the skyrmion}

We already know the origin of the asymmetry of the magnetization texture in the stripe, let us analyze how in turn, the imprint affects the skyrmion, and how the subsequent relaxation process of these mutually interacting magnetic subsystems proceeds. To get a deeper insight into a mechanism of the skyrmion deformation, we performed independent simulations for the subunit of the structure, it is a stripe with a frozen magnetic texture, i.e., a skyrmion imprint present, but a nanodot removed.

Figure~\ref{fig:sub-units}(b) shows $\textbf{H}_{\textrm{s-str}}(x,y,z=6.5 \text{ nm})$, i.e., the magnetostatic stray filed from the imprint in the plane between the dot and the stripe, in Supplementary Material Fig.~\ref{fig:SUP:b_demag_without_nd_zcross3} this field along the plane perpendicular to the stripe axis is shown. This stray field is opposite to and an order of magnitude lower than the field, which is created by the nanodot $\textbf{H}_{\textrm{s-dot}}$ [compare Fig.~\ref{fig:sub-units} (a)], with maximum stray field intensity $\Mu_{0} H_{\textrm{s-str}} = 0.036$ T. Comparing the imprint for positive and negative values of DMI [see Fig.~\ref{fig:images_dmi_depenece}(b-c)], we can see that the weaker field induced by the skyrmion for positive DMI induces a weaker imprint, which in turn generates weaker $\textbf{H}_{\textrm{s-str}}$. Moreover, as shown in the previous sub-section a strength of the imprint, and so the $\textbf{H}_{\textrm{s-str}}$, depends also on the rigidity of the magnetization in the stripe, which is determined by the magnetic anisotropy.
This clearly indicates that the amplitude of $\textbf{H}_{\textrm{s-str}}$ field is proportional to the local deformation of the imprint. Thus, the larger imprint, the stronger $y$-component field at the skyrmion-domain wall position is, and it is oriented toward the $x$-axis [see, Fig.~\ref{fig:sub-units}(b)]. This stray-field component influences the skyrmion resulting in its squeezing along the $y$-axis, and finally deformation of the circular skyrmion to the elliptical shape. Moreover, due to asymmetric imprint, the stronger $\textbf{H}_{\textrm{s-str}}$ field is for $x<0$ (under assumed geometry), which additionally deforms the ellipsoidal shape towards the egg-shape skyrmion as seen in Figs.~\ref{fig:single_static}-\ref{fig:images_dmi_depenece}. 

To demonstrate the impact of the internal magnetic field in the stripe on the skyrmion shape in the nanodot, we performed simulations in which we assumed uniaxial (parallel to the $x$ axis) magnetocrystalline anisotropy in the Py stripe with different values of the anisotropy constant $K_\mathrm{u}$. Fig.~\ref{fig:sub-units}(d) presents simulation results for DMI value -1.0 mJ/m$^{2}$, where the increase of the anisotropy clearly results in the decrease of skyrmion size and the skyrmion asymmetry. 
Starting with 0 magnetocrystalline anisotropy [the egg-shaped skyrmion of 200 nm maximal size, see vertical light-blue dashed line in Fig.~\ref{fig:d_dependence}(a)] the skyrmion size and its asymmetry rapidly decrease with increasing anisotropy up to $K_\mathrm{u} \approx 10$ kJ/m$^3$, above it reaches the circular shape. Further, an increase in anisotropy constant results in a slow monotonous reduction of the skyrmion size, and for $K_\mathrm{u} = 30$ kJ/m$^3$ the skyrmion has only 25 nm in diameter. We presume that at infinite large anisotropy, the skyrmion reaches a size of 17 nm, i.e., the size of the skyrmion in the isolated nanodot. Thus, by controlling the strength of the internal magnetic field in the stripe, e.g., via uniaxial anisotropy or the external magnetic field, we can control the degree of deformation and the size of the skyrmion, and so control the effective role of the DMI in the nanodot. 

The last unexplained effect observed in simulations in Figs.~\ref{fig:images_dmi_depenece}-\ref{fig:d_dependence} is a significant enlarging of the skyrmion size in the hybrid structure with respect to the isolated dot, above DMI value -0.88 and 1.16 mJ/m$^2$, for negative and positive DMI, respectively. We attribute this effect to the reduction of the internal magnetic field, in particular its $z$ component, inside the nanodot. This is due to the compensation of the magnetostatic charges on the nanodot surface by the surface charges generated on the ferromagnetic stripe by the out-of-plane component of the imprint magnetization, which is always parallel to the polarization of the nanodot domains, see Fig.~\ref{fig:SUP_sk_polarization} in Supplemental Materials. Thus, the influence of the surface charges from the imprint can be similar to the role of the external magnetic field parallel to the skyrmion core. 
Finally, we can speculate that increasing the magnetization saturation of the nanodot, in additionally to decreasing  the anisotropy of the stripe (discussed in the previous paragraph) shall allow to increase in the size of the skyrmion, and increase the DMI range of large skyrmion stability. Note that the described processes proceed at a mutual relaxation of the two magnetization textures, deepening the imprint and deformation of the skyrmion. Our discussion focuses on the analysis of magnetostatic forces for different values of DMI, but the degree of deformation is also affected by the strength of exchange interactions or effective anisotropy in the nanodot as well as in the stripe. Nevertheless, even though the relaxation process in 3D hybrid structures is complex, the main effects can be elucidated from the analysis of stray filed distributions, as we have just shown. 

\subsection*{Application outlook}

In the last section, we examine the movement of the skyrmion and the imprint in the hybrid system driven by the spin-trasfer torque (STT) generated by an electric current applied to ferromagnetic layers. To do that, we replace the Pt/Co/Ir nanodot with a respective stripe of the 300 nm width, placed 1.5 nm above an extended Py film of 4.5 nm thickness. To preserve the effective anisotropy maintaining uniform magnetization in the plane of the Py layer, we assumed a uniaxial anisotropy perpendicular to the Pt/Co/Ir stripe axis with $K_{\mathrm{u}}=3.5$ kJ/m$^{3}$. 

\begin{figure}[!htp]
 \includegraphics[width=\textwidth]{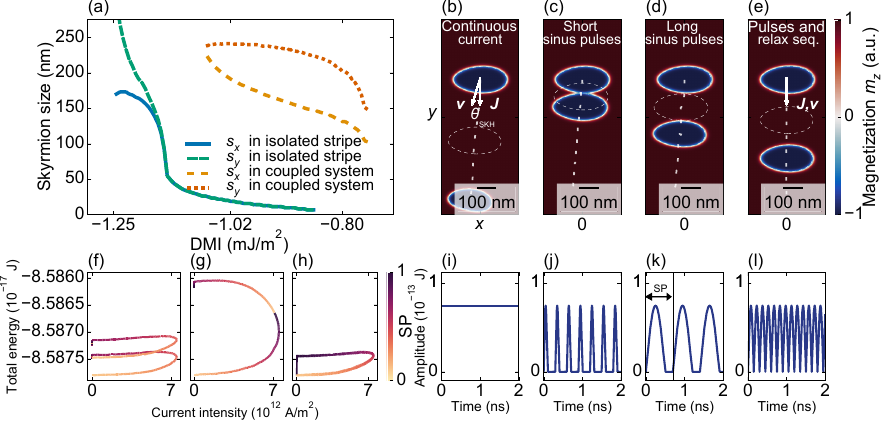}
 \caption{(a) The skyrmion size dependence on DMI strength for the skyrmion in an isolated stripe and the skyrmion in the stripe coupled with the Py layer. (b-e) Simulated skyrmion positions driven by STT for four different scenarios of the current and relaxation time sequences performed for DMI = 0.8 mJ/m$^{2}$: (b) continuous electric current, (c) 0.1 ns electric current pulse with the following 0.2 ns relaxation, (d) 0.5 ns current pulse and 0.2 ns relaxation. The relaxation process in simulations in (b-d) was performed in real-time (RT). (e) 0.1 ns current pulse and time-less relaxation. In these images are present the initial and final, after 2 ns, skyrmion position. White-dashed contours present skyrmion position after 1 ns. The straight dashed line presents the skyrmion trajectory. Panels (f-h) present simulated total energy loops in dependence on the applied current intensity for the three different time sequence scenarios presented on panels (c-e), respectively. Color points at the time evolution in one sequence period (SP) of the electric current change [see panels (j-l)]. Panels (i-l) present time-dependent current intensity profiles used for simulations shown in (b-e), respectively. 
 }
 \label{fig:racetrack_1}
\end{figure}

To determine the DMI range of the skyrmion stability in this new geometry, we examine the dependence of skyrmion size on the DMI strength, the results are shown in Fig.~\ref{fig:racetrack_1}(a). For a reference system, i.e., an isolated Pt/Co/Ir stripe, we found circular skyrmions from $-0.86$ mJ/m$^{2}$ to $-1.17$ mJ/m$^{2}$. Above this value, the skyrmion begins to grow anisotropically toward the long axis of the stripe until the value of $-1.25$ mJ/m$^{2}$, where the skyrmion size is $s_x = 172$~nm and $s_y = 272$~nm, along the $x$ and $y$ axis, respectively. For higher DMI values, skyrmion evolves to complex magnetic textures. 

For the hybrid system, we obtain a stable skyrmion for lower DMI values compared to the isolated stripe. Analogously to the nanodot over Py stripe, a magnetostatic coupling causes skyrmion deformation and increases its size. For DMI 0.733 mJ/m$^{2}$ we found a skyrmion with the size of $s_x=148.6$ nm, and $s_y=102.4$ nm. The upper limit of stability is difficult to determine because above 1.01 mJ/m$^{2}$ a skyrmion deformation disrupts the regular shape of the skyrmion. For a DMI of 1.032 mJ/m$^{2}$, the assumed skyrmion stability limit, we observe a skyrmion with dimensions $s_x =$ 220 nm and $s_y = 189$ nm. The micromagnetic simulation results performed for DMI $=0.8$ mJ/m$^{2}$ presented in Fig.~\ref{fig:racetrack_1}(b) show that the application of a 2 ns continuous electric current [a current signal is shown in Fig.~\ref{fig:racetrack_1}(i)] with the amplitude of 10$^{13}$ J/m$^2$ leads to the move of the skyrmion but with deflection of its trajectory and its squeezing. After 4 ns it moved for a distance of about 3 $\Mu$m with a velocity of around 414 m/s. Interestingly, after the current is turned off the skyrmion relaxes to the initial state, regardless of the position in the stripe. This observation allows us to propose a scenario for skyrmion transport along the long track by current pulses interspersed with skyrmion relaxation periods.

In Fig.~\ref{fig:racetrack_1}(e) we demonstrate the concept of such a scenario in which each pulse of the electric current is followed by a full relaxation process. The white dashed line in Fig.~\ref{fig:racetrack_1}(e) shows the straight trajectory of the motion in such a sequence, where the SHE during the current pulse is fully compensated during the relaxation. Panel (h) shows closed loops of the total energy of the system dependent on the applied current, which remains constant for each sequence period (SP) and preserves a skyrmion position across the stripe. To verify our concept, we applied the electric current along the $y$ axis in a form of a sin function:
\begin{equation}
 J_{y}(t) =
 \begin{cases}
 \textit{J}_{0}\sin(\pi \frac{t}{T_{\mathrm{c}}}) &
 \text{if $t \in (0 ; T_{\mathrm{c}}) $},\\
 0 & \text{if $t \in (T_{\mathrm{c}};T_{\mathrm{r}}) $},
 \end{cases},
\end{equation}
where $\textit{J}_0$ is a current density amplitude $10^{13}$ A/m$^2$, $T_{\textrm{c}}$ is a period of applied current forming together with the relaxation period $T_{\textrm{r}}$ a sequence period ($\textrm{SP}=T_{\textrm{c}}+T_{\textrm{r}}$). The simulation results for different SPs are shown in Fig.~\ref{fig:racetrack_1}(c-d) and their respective time courses in panels (j-k), for the two values of $T_{\textrm{c}}$, i.e. 0.1 ns and 0.5 ns. For both cases, the $T_{\textrm{r}}$ was 0.2 ns. In the first case, the skyrmion moved 194 nm in 2 ns at the velocity 97 m/s, in the second case, it moved 374 nm at 187 m/s. In the first case, after 2 ns of moving, the skyrmion has shrunk by about 8 nm along the short axis and about 3 nm along the long axis of the stripe with respect to the original size, with the translation to the left around 12 nm. During the relaxation time, the skyrmion has increased by 3 and 1 nm, along the short and long axis of the stripe, respectively, without the change in the position along the stripe. The assumed relaxation time does not allow the skyrmion to return to its original size, so with further repetitions of the current pulses, the skyrmion underwent deepening deformation and shrinkage. For the longer field pulse, the deformation and the shift toward the edge were even larger (with $T_{\textrm{c}} = 0.5$ ns and $T_{\textrm{r}}=0.2$ ns, the skyrmion has shrunk by about 14 nm in the short axis and about 5 nm in the long axis of the stripe, relative to the original dimensions, with the translation to the left around 32 nm). 

The skyrmion transport process can be expressed also as a total energy function in dependence on the applied current density and time. It is presented in Fig.~\ref{fig:racetrack_1}(f-h) for different sequence procedures presented in Fig.~\ref{fig:racetrack_1}(c-e), respectively. For instance, Fig.~\ref{fig:racetrack_1}(f) represents the total energy during the first two SPs. Notice that as the current intensity increases above $4\cdot10^{12}$ A/m$^2$, the skyrmion begins to move. With further current intensity increase, the skyrmion continues to move, and its energy increases until the current intensity is reduced below $4\cdot10^{12}$ A/m$^2$. The relaxation stage is visible as a vertical line at 0 value of the current intensity, during which a falling down of the total energy of the system is observed. At this time there is a slow stretching of the skyrmion and a return to the original dimensions. In both cases (f) and (g), a time of 0.2 ns is insufficient to fully relax the system. For a longer electrical pulse $T_{\text{c}}$, it can be seen that the total energy has increased significantly, but the system returns to the ground state faster ($1.5\cdot10^{-4}$ A/m in comparison to $8\cdot10^{-4}$ A/m in the case of shorter SP). 
Our results show that a technique for unconstrained skyrmion transport along the race-track based on the hybrid systems is possible but requires further optimization. In particular, the crucial is establishing the duration of the electric current pulse relative to the relaxation time, because an extending the duration of the electric pulse, the skyrmion travels a longer distance but is more severely deformed. The needed relaxation time depends on many factors such as material parameters, damping, and temperature, additionally to an electric current pulse length. However, such considerations need separate consideration.

\subsection*{Conclusions}

Using micromagnetic simulations we described the skyrmion's symmetry-breaking mutual magnetostatic interactions in the system composed of the nanodot, possessing a N\'eel skyrmion and the soft ferromagnetic stripe. 
In this system, the skyrmion exerts a stray magnetostatic field that deforms the magnetization in the stripe. Consequently, the magnetization texture imprinted upon the stripe exerts a stray field on the skyrmion, causing its enlarging and deforming into an egg-like shape. We found that in this hybrid system the size of the skyrmion is much more sensitive to the changes in DMI value than in the isolated dot. We demonstrated that here the DMI range of skyrmion stability is extended, allowing relatively small skyrmions to be stabilized at relatively low DMI values. Additionally, we found that for some range of DMI values and configuration, the interaction with the stripe creates suitable conditions for bi-stability of the skyrmion state, corresponding to a large and a small skyrmion, it is with the size above 110 nm and below 40 nm, respectively, in our hybrid structure case.

We show that the skyrmion and its deformation in hybrid structures of this type can be controlled by internal and external parameters (see graphical summary in Fig.~S2, Supl. Matter). In particular, the polarization of the skyrmion core does not affect the strength of the interactions between the skyrmion and the stripe, but it controls the asymmetry orientation of the skyrmion with respect to the plane perpendicular to the axis of the stripe. An analogous changes can be achieved by reversing the orientation of the magnetization in the stripe. 
The sign of the DMI determines the chirality of the N{\'e}el skyrmion and so the side of the nanodot with the enhanced stray magnetostatic field. This Halbach effect in nanoscale can be used to increase or decrease the mutual interaction of the nanodot with the stripe, just by the change of the order deposition. Finally, the skyrmion size and deformation strength can be controlled by the internal magnetic field of the stripe, in particular by the effective magnetic anisotropy or bias magnetic field. 

The skyrmion-imprint coupled magnetization structure in the hybrid system can be used to overcome the skyrmion Hall effect undesirable in the application context, and is suitable to transport a skyrmion along the straight track by electric current pulses. We have presented a proof-of-concept technique for unconstrained transport of skyrmions along a racetrack composed of a hybrid ferromagnetic system. Our results allow us to understand the magnetostatic interactions in chiral 3D structures, enabling better design of skyrmionic and magnonic devices. 

\section*{Methods}

The micromagnetic simulations are performed by using the Mumax3\cite{Leliaert2018FastMumax3} which solves the Landau-Lifshitz-Gilbert equation:
\begin{equation}
 \frac{\text{d}\mathbf{m}}{\mathrm{d}t}=\gamma \Mu_0 \frac{1}{1+\alpha^{2}} (\mathbf{m} \times  \mathbf{H}_{\mathrm{eff}}) + \alpha\Mu_0 \left( \mathbf{m} \times (\mathbf{m} \times \mathbf{H}_{\mathrm{eff}}) \right),
\end{equation}
where $\textbf{m} = \textbf{M} / \mathrm{M}_{\mathrm{s}}$ is the normalized magnetization, $\textbf{\text{H}}_{\mathrm{eff}}$ is the effective magnetic field acting on the magnetization, $\gamma=-1.7595 \cdot 10^{11}$ Hz/T is the gyromagnetic ration, $\alpha$ is the Gilbert damping. 
In this paper, the following components were considered for the effective field $\textbf{H}_{\mathrm{eff}}$: demagnetizing field $\textbf{\text{H}}_{\mathrm{d}}$, exchange field $\textbf{\text{H}}_{\mathrm{exch}}$,
Dzyaloshinskii-Moriya exchange field $\textbf{\text{H}}_{\mathrm{DMI}}$, and magneto-crystalline anisotropy field $\textbf{\text{H}}_{\mathrm{Ku}}$. External maagnetic field and thermal effects were neglected. Thus, the effective field term $\textbf{H}_{\mathrm{eff}}$ is expressed as:
\begin{equation}
    \textbf{H}_{\mathrm{eff}} =
    \textbf{H}_{\mathrm{d}} + \textbf{H}_{\mathrm{exch}} + \textbf{H}_{\mathrm{DMI}} + \textbf{H}_{\mathrm{Ku}},
\end{equation}
where
\begin{equation}
    \textbf{H}_{\mathrm{exch}} = 2 \frac{A_{\mathrm{ex}}}{\Mu_0 \mathrm{M}_{\mathrm{s}}} \Delta \textbf{m},
\end{equation}
\begin{equation}
\textbf{H}_{\mathrm{DMI}} = \frac{2 \mathrm{DMI}}{\Mu_0\mathrm{M}_{\mathrm{s}}} 
\left(\frac{\partial m_z}{\partial x},\frac{\partial m_z}{\partial y},-\frac{\partial m_x}{\partial x},-\frac{\partial m_y}{\partial y}, \right),
\end{equation}
and the uniaxial magneto-crystalline anisotropy in the form:
\begin{equation}
\textbf{H}_{\mathrm{Ku}} =
\frac{2\textbf{K}_{\mathrm{u}}}{\Mu_0 \textrm{M}_{\mathrm{s}}} 
\left( \textbf{u} \cdot \textbf{m} \right) \textbf{u},
\end{equation}
where $\textbf{K}_{\mathrm{u}}$ is the first order uniaxial anisotropy constant and $\textbf{u}$ is an unit vector indicating the anisotropy direction.

Throughout the paper, for a nanodot we use following material parameters: saturation magnetization M$_\mathrm{s}$ = 956 kA/m, exchange stiffness constant A$_{\text{ex}} = 10$ pJ/m, DMI constant in a ragne from 0 to 2 mJ/m$^2$, out-of-plane magnetic anisotropy constant $K_{\mathrm{u}} = 0.717$ MJ/m$^3$ and Gilbert damping constant $\mathrm{\alpha} =$ 0.03. This set of parameters corresponds to the ultrathin layer of Pt/Co/Ir~\cite{PSSR:PSSR201700259,Moreau-Luchaire2016AdditiveTemperature} which presents favorable conditions for the skyrmion stabilization. The parameters used to model the stripe correspond to a 4.5 nm thick layer of Py, and has the following magnetic parameters: saturation magnetization $M_\mathrm{s} = 800$ kA/m, exchange stiffness constant $A_{\mathrm{ex}} = 13$~pJ/m and Gilbert damping constant $\mathrm{\alpha} = 0.01$. Additionally, we assumed absorbing boundary conditions at both horizontal sides of the stripe. The studied system was discretized uniformly with a 0.75 $\times$ 0.75 $\times$ 1.5 nm$^{3}$ unit cells. 

The relaxation simulations of the hybridized system we performed using a sequence of Mumax built-in relaxation functions in four steps: minimization (\textit{minimize}), relaxation (\textit{relax}) with default energy threshold values, solving the full LLG equation (simulation run for 1 ns), and energy minimization again. This sequence allowed us to reduce the total energy and obtain stable magnetization configurations.
The bi-stable skyrmion states where computed by performing simulations of skyrmion stabilization as a function of DMI intensity, assuming the two skyrmion relaxations, one when the initial state is a large-diameter skyrmion (75\% of the disk diameter) and the second with a small-diameter skyrmion (10\% of the disk diameter). 

We used the frozen-spin technique (using a \textit{frozen} function of Mumax) to extract the 
spatial distribution of the magnetostatic fields induced only from a deformed skyrmion or an imprint, without the presence of a second  magnetic element. We calculated these distributions by freezing the magnetic texture after relaxation and simultaneously removing the second element, then by performing one simulation step, solver automatically calculates the magnetostatic field distributions. The sum of field distributions from both elements is equal to the field intensity obtained for the coupled system. 

In order to model the influence of the uniform electric current density on the skyrmion  we use the Zhang-Li extension to the LLG equation~\cite{mumax_2014,Zhang2004}:
\begin{align*}
 \frac{\text{d}\mathbf{M}}{\mathrm{d}t}= &\gamma \Mu_0 \frac{1}{1+\alpha^{2}} (\mathbf{m} \times  \mathbf{H}_{\mathrm{eff}}) + \alpha\Mu_0 \left( \mathbf{m} \times (\mathbf{m} \times \mathbf{H}_{\mathrm{eff}}) \right)\\
 &- \frac{\nu}{\text{m}^{2}_{\text{s}}}\mathbf{m} \times \left(\mathbf{m} \times \hat{j} \cdot \nabla \textbf{m} \right) - 
  \frac{\xi \nu}{\text{M}_{\text{s}}}\left(\mathbf{m} \times \hat{j} \cdot \nabla \textbf{m} \right),
\end{align*}
The current density appears only throughout the quantity $\nu$, which is defined as:
\begin{equation}
    \nu = \frac{P j \Mu_{B}}
    { e \textbf{M}_{\mathrm{s}}(1+{\xi^{2}})},
\end{equation}
where $\hat{j}$ is the unit vector defining the direction of the electric current flow, $P$ is the degree of polarization of the spin current, $\xi$ is degree of non adiabadicity, $\Mu_{B}$ is Bohr magneton, $e$ is the electron charge. In our Mumax simulations regarding racetrack memories we assumed positive polarization ($P=1$) and neglected parameter $\xi$, we also assumed the uniform spatial distribution of the current density. We considered a uniform electric current density distribution with an amplitude of $1\cdot 10^{13}$ J/m$^2$. 
For race-track simulations we used Runge-Kutta RK45 solver with the following time-step parameters $\texttt{maxerr}=1\cdot 10^{-5}$, \texttt{mindt} =$ 1.0\cdot 10^{-14}$, \texttt{maxdt} = $3.0\cdot 10^{-14}$. We assumed 32 repetitions along the $x$-axis and one along the $y$-axis in the periodic boundary conditions. These simulations were performed for Py rectangular stripe of 1200 $ \times $ 975 $ \times $ 5 nm$^3$ using an uniformly discretized grid with a cell size of 0.75 $ \times $ 0.75 $ \times $ 1.5 nm$^{3}$ along the $x,y$, and $z$ axes, respectively. The width of the reca-track is 300 nm and its length is 975 nm. 

\begin{acknowledgments}
The authors acknowledge the fruitful discussions with prof. Jaros\l{}aw W. K\l{}os. The research has received funding from National Science Centre of Poland, Grant No.~UMO-2018/30/Q/ST3/00416. 
M. M. acknowledges funding from the Slovak Grant Agency APVV (grant number APVV-19-0311(RSWFA)), and ITMS, project code 313021T081 supported by the Research and Innovation Operational Programme funded by the ERDF. The simulations were partially performed at the Poznan Supercomputing and Networking Center (Grant No.~398). 
\end{acknowledgments}

\bibliography{manusctipy.bbl}

\providecommand{\latin}[1]{#1}
\makeatletter
\providecommand{\doi}
  {\begingroup\let\do\@makeother\dospecials
  \catcode`\{=1 \catcode`\}=2 \doi@aux}
\providecommand{\doi@aux}[1]{\endgroup\texttt{#1}}
\makeatother
\providecommand*\mcitethebibliography{\thebibliography}
\csname @ifundefined\endcsname{endmcitethebibliography}
  {\let\endmcitethebibliography\endthebibliography}{}
\begin{thebibliography}{40}%
\makeatletter
\providecommand \@ifxundefined [1]{%
 \@ifx{#1\undefined}
}%
\providecommand \@ifnum [1]{%
 \ifnum #1\expandafter \@firstoftwo
 \else \expandafter \@secondoftwo
 \fi
}%
\providecommand \@ifx [1]{%
 \ifx #1\expandafter \@firstoftwo
 \else \expandafter \@secondoftwo
 \fi
}%
\providecommand \natexlab [1]{#1}%
\providecommand \enquote  [1]{``#1''}%
\providecommand \bibnamefont  [1]{#1}%
\providecommand \bibfnamefont [1]{#1}%
\providecommand \citenamefont [1]{#1}%
\providecommand \href@noop [0]{\@secondoftwo}%
\providecommand \href [0]{\begingroup \@sanitize@url \@href}%
\providecommand \@href[1]{\@@startlink{#1}\@@href}%
\providecommand \@@href[1]{\endgroup#1\@@endlink}%
\providecommand \@sanitize@url [0]{\catcode `\\12\catcode `\$12\catcode
  `\&12\catcode `\#12\catcode `\^12\catcode `\_12\catcode `\%12\relax}%
\providecommand \@@startlink[1]{}%
\providecommand \@@endlink[0]{}%
\providecommand \url  [0]{\begingroup\@sanitize@url \@url }%
\providecommand \@url [1]{\endgroup\@href {#1}{\urlprefix }}%
\providecommand \urlprefix  [0]{URL }%
\providecommand \Eprint [0]{\href }%
\providecommand \doibase [0]{https://doi.org/}%
\providecommand \selectlanguage [0]{\@gobble}%
\providecommand \bibinfo  [0]{\@secondoftwo}%
\providecommand \bibfield  [0]{\@secondoftwo}%
\providecommand \translation [1]{[#1]}%
\providecommand \BibitemOpen [0]{}%
\providecommand \bibitemStop [0]{}%
\providecommand \bibitemNoStop [0]{.\EOS\space}%
\providecommand \EOS [0]{\spacefactor3000\relax}%
\providecommand \BibitemShut  [1]{\csname bibitem#1\endcsname}%
\let\auto@bib@innerbib\@empty
\bibitem [{\citenamefont {Moreau-Luchaire}\ \emph {et~al.}(2016)\citenamefont
  {Moreau-Luchaire}, \citenamefont {Moutafis}, \citenamefont {Reyren},
  \citenamefont {Sampaio}, \citenamefont {Vaz}, \citenamefont {Van~Horne},
  \citenamefont {Bouzehouane}, \citenamefont {Garcia}, \citenamefont
  {Deranlot}, \citenamefont {Warnicke}, \citenamefont {Wohlh{\"{u}}ter},
  \citenamefont {George}, \citenamefont {Weigand}, \citenamefont {Raabe},
  \citenamefont {Cros},\ and\ \citenamefont
  {Fert}}]{Moreau-Luchaire2016AdditiveTemperature}%
  \BibitemOpen
  \bibfield  {author} {\bibinfo {author} {\bibfnamefont {C.}~\bibnamefont
  {Moreau-Luchaire}}, \bibinfo {author} {\bibfnamefont {C.}~\bibnamefont
  {Moutafis}}, \bibinfo {author} {\bibfnamefont {N.}~\bibnamefont {Reyren}},
  \bibinfo {author} {\bibfnamefont {J.}~\bibnamefont {Sampaio}}, \bibinfo
  {author} {\bibfnamefont {C.~A.}\ \bibnamefont {Vaz}}, \bibinfo {author}
  {\bibfnamefont {N.}~\bibnamefont {Van~Horne}}, \bibinfo {author}
  {\bibfnamefont {K.}~\bibnamefont {Bouzehouane}}, \bibinfo {author}
  {\bibfnamefont {K.}~\bibnamefont {Garcia}}, \bibinfo {author} {\bibfnamefont
  {C.}~\bibnamefont {Deranlot}}, \bibinfo {author} {\bibfnamefont
  {P.}~\bibnamefont {Warnicke}}, \bibinfo {author} {\bibfnamefont
  {P.}~\bibnamefont {Wohlh{\"{u}}ter}}, \bibinfo {author} {\bibfnamefont
  {J.~M.}\ \bibnamefont {George}}, \bibinfo {author} {\bibfnamefont
  {M.}~\bibnamefont {Weigand}}, \bibinfo {author} {\bibfnamefont
  {J.}~\bibnamefont {Raabe}}, \bibinfo {author} {\bibfnamefont
  {V.}~\bibnamefont {Cros}},\ and\ \bibinfo {author} {\bibfnamefont
  {A.}~\bibnamefont {Fert}},\ }\bibfield  {title} {\bibinfo {title} {{Additive
  interfacial chiral interaction in multilayers for stabilization of small
  individual skyrmions at room temperature}},\ }\href
  {https://doi.org/10.1038/nnano.2015.313} {\bibfield  {journal} {\bibinfo
  {journal} {Nature Nanotechnology}\ }\textbf {\bibinfo {volume} {11}},\
  \bibinfo {pages} {444} (\bibinfo {year} {2016})}\BibitemShut {NoStop}%
\bibitem [{\citenamefont {Marrows}(2015)}]{Marrows2015}%
  \BibitemOpen
  \bibfield  {author} {\bibinfo {author} {\bibfnamefont {C.}~\bibnamefont
  {Marrows}},\ }\bibfield  {title} {\bibinfo {title} {{An inside view of
  magnetic skyrmions}},\ }\href {https://doi.org/10.1103/Physics.8.40}
  {\bibfield  {journal} {\bibinfo  {journal} {Physics}\ }\textbf {\bibinfo
  {volume} {8}},\ \bibinfo {pages} {40} (\bibinfo {year} {2015})}\BibitemShut
  {NoStop}%
\bibitem [{\citenamefont {Kiselev}\ \emph {et~al.}(2011)\citenamefont
  {Kiselev}, \citenamefont {Bogdanov}, \citenamefont {Sch{\"{a}}fer},\ and\
  \citenamefont {R{\"{o}}{\ss}ler}}]{Kiselev_chilar_skyrmions}%
  \BibitemOpen
  \bibfield  {author} {\bibinfo {author} {\bibfnamefont {N.~S.}\ \bibnamefont
  {Kiselev}}, \bibinfo {author} {\bibfnamefont {a.~N.}\ \bibnamefont
  {Bogdanov}}, \bibinfo {author} {\bibfnamefont {R.}~\bibnamefont
  {Sch{\"{a}}fer}},\ and\ \bibinfo {author} {\bibfnamefont {U.~K.}\
  \bibnamefont {R{\"{o}}{\ss}ler}},\ }\bibfield  {title} {\bibinfo {title}
  {{Chiral skyrmions in thin magnetic films: new objects for magnetic storage
  technologies?}},\ }\href {https://doi.org/10.1088/0022-3727/44/39/392001}
  {\bibfield  {journal} {\bibinfo  {journal} {Journal of Physics D: Applied
  Physics}\ }\textbf {\bibinfo {volume} {44}},\ \bibinfo {pages} {392001}
  (\bibinfo {year} {2011})}\BibitemShut {NoStop}%
\bibitem [{\citenamefont {Zhang}\ \emph
  {et~al.}(2020{\natexlab{a}})\citenamefont {Zhang}, \citenamefont {Zhou},
  \citenamefont {Mee~Song}, \citenamefont {Park}, \citenamefont {Xia},
  \citenamefont {Ezawa}, \citenamefont {Liu}, \citenamefont {Zhao},
  \citenamefont {Zhao},\ and\ \citenamefont
  {Woo}}]{Zhang2020Skyrmion-electronics:Applications}%
  \BibitemOpen
  \bibfield  {author} {\bibinfo {author} {\bibfnamefont {X.}~\bibnamefont
  {Zhang}}, \bibinfo {author} {\bibfnamefont {Y.}~\bibnamefont {Zhou}},
  \bibinfo {author} {\bibfnamefont {K.}~\bibnamefont {Mee~Song}}, \bibinfo
  {author} {\bibfnamefont {T.-E.}\ \bibnamefont {Park}}, \bibinfo {author}
  {\bibfnamefont {J.}~\bibnamefont {Xia}}, \bibinfo {author} {\bibfnamefont
  {M.}~\bibnamefont {Ezawa}}, \bibinfo {author} {\bibfnamefont
  {X.}~\bibnamefont {Liu}}, \bibinfo {author} {\bibfnamefont {W.}~\bibnamefont
  {Zhao}}, \bibinfo {author} {\bibfnamefont {G.}~\bibnamefont {Zhao}},\ and\
  \bibinfo {author} {\bibfnamefont {S.}~\bibnamefont {Woo}},\ }\bibfield
  {title} {\bibinfo {title} {{Skyrmion-electronics: writing, deleting, reading
  and processing magnetic skyrmions toward spintronic applications}},\ }\href
  {https://doi.org/10.1088/1361-648X/ab5488} {\bibfield  {journal} {\bibinfo
  {journal} {Journal of Physics: Condensed Matter}\ }\textbf {\bibinfo {volume}
  {32}},\ \bibinfo {pages} {143001} (\bibinfo {year}
  {2020}{\natexlab{a}})}\BibitemShut {NoStop}%
\bibitem [{\citenamefont {Zhou}(2019)}]{Zhou2019MagneticConcepts}%
  \BibitemOpen
  \bibfield  {author} {\bibinfo {author} {\bibfnamefont {Y.}~\bibnamefont
  {Zhou}},\ }\bibfield  {title} {\bibinfo {title} {{Magnetic skyrmions:
  intriguing physics and new spintronic device concepts}},\ }\href
  {https://doi.org/10.1093/nsr/nwy109} {\bibfield  {journal} {\bibinfo
  {journal} {National Science Review}\ }\textbf {\bibinfo {volume} {6}},\
  \bibinfo {pages} {210} (\bibinfo {year} {2019})}\BibitemShut {NoStop}%
\bibitem [{\citenamefont {Finocchio}\ and\ \citenamefont
  {Panagopoulos}(2021)}]{Finocchio2021MagneticApplications}%
  \BibitemOpen
  \bibfield  {author} {\bibinfo {author} {\bibfnamefont {G.}~\bibnamefont
  {Finocchio}}\ and\ \bibinfo {author} {\bibfnamefont {C.}~\bibnamefont
  {Panagopoulos}},\ }\href {https://doi.org/10.1016/C2019-0-02206-6} {\emph
  {\bibinfo {title} {Magnetic Skyrmions and Their Applications}}}\ (\bibinfo
  {publisher} {Woodhead Publishing Series in Electronic and Optical Materials,
  Elsevier},\ \bibinfo {year} {2021})\BibitemShut {NoStop}%
\bibitem [{\citenamefont {Wang}\ \emph {et~al.}(2018)\citenamefont {Wang},
  \citenamefont {Yuan},\ and\ \citenamefont {Wang}}]{Wang2018ASize}%
  \BibitemOpen
  \bibfield  {author} {\bibinfo {author} {\bibfnamefont {X.~S.}\ \bibnamefont
  {Wang}}, \bibinfo {author} {\bibfnamefont {H.~Y.}\ \bibnamefont {Yuan}},\
  and\ \bibinfo {author} {\bibfnamefont {X.~R.}\ \bibnamefont {Wang}},\
  }\bibfield  {title} {\bibinfo {title} {{A theory on skyrmion size}},\ }\href
  {https://doi.org/10.1038/s42005-018-0029-0} {\bibfield  {journal} {\bibinfo
  {journal} {Communications Physics}\ }\textbf {\bibinfo {volume} {1}},\
  \bibinfo {pages} {31} (\bibinfo {year} {2018})}\BibitemShut {NoStop}%
\bibitem [{\citenamefont {Yang}\ \emph {et~al.}(2015)\citenamefont {Yang},
  \citenamefont {Thiaville}, \citenamefont {Rohart}, \citenamefont {Fert},\
  and\ \citenamefont {Chshiev}}]{Yang2015AnatomyInterfaces}%
  \BibitemOpen
  \bibfield  {author} {\bibinfo {author} {\bibfnamefont {H.}~\bibnamefont
  {Yang}}, \bibinfo {author} {\bibfnamefont {A.}~\bibnamefont {Thiaville}},
  \bibinfo {author} {\bibfnamefont {S.}~\bibnamefont {Rohart}}, \bibinfo
  {author} {\bibfnamefont {A.}~\bibnamefont {Fert}},\ and\ \bibinfo {author}
  {\bibfnamefont {M.}~\bibnamefont {Chshiev}},\ }\bibfield  {title} {\bibinfo
  {title} {{Anatomy of Dzyaloshinskii-Moriya Interaction at Co/Pt
  Interfaces}},\ }\href {https://doi.org/10.1103/PhysRevLett.115.267210}
  {\bibfield  {journal} {\bibinfo  {journal} {Physical Review Letters}\
  }\textbf {\bibinfo {volume} {115}},\ \bibinfo {pages} {267210} (\bibinfo
  {year} {2015})}\BibitemShut {NoStop}%
\bibitem [{\citenamefont {Behera}\ \emph {et~al.}(2018)\citenamefont {Behera},
  \citenamefont {Mishra}, \citenamefont {Mallick}, \citenamefont {Singh},\ and\
  \citenamefont {Bedanta}}]{Behera2018SizeAnisotropy}%
  \BibitemOpen
  \bibfield  {author} {\bibinfo {author} {\bibfnamefont {A.~K.}\ \bibnamefont
  {Behera}}, \bibinfo {author} {\bibfnamefont {S.~S.}\ \bibnamefont {Mishra}},
  \bibinfo {author} {\bibfnamefont {S.}~\bibnamefont {Mallick}}, \bibinfo
  {author} {\bibfnamefont {B.~B.}\ \bibnamefont {Singh}},\ and\ \bibinfo
  {author} {\bibfnamefont {S.}~\bibnamefont {Bedanta}},\ }\bibfield  {title}
  {\bibinfo {title} {{Size and shape of skyrmions for variable
  Dzyaloshinskii-Moriya interaction and uniaxial anisotropy}},\ }\href
  {https://doi.org/10.1088/1361-6463/aac9a7} {\bibfield  {journal} {\bibinfo
  {journal} {Journal of Physics D: Applied Physics}\ }\textbf {\bibinfo
  {volume} {51}},\ \bibinfo {pages} {285001} (\bibinfo {year}
  {2018})}\BibitemShut {NoStop}%
\bibitem [{\citenamefont {Zhang}\ \emph
  {et~al.}(2016{\natexlab{a}})\citenamefont {Zhang}, \citenamefont {Xia},
  \citenamefont {Zhou}, \citenamefont {Wang}, \citenamefont {Liu},
  \citenamefont {Zhao},\ and\ \citenamefont {Ezawa}}]{Zhang2016b}%
  \BibitemOpen
  \bibfield  {author} {\bibinfo {author} {\bibfnamefont {X.}~\bibnamefont
  {Zhang}}, \bibinfo {author} {\bibfnamefont {J.}~\bibnamefont {Xia}}, \bibinfo
  {author} {\bibfnamefont {Y.}~\bibnamefont {Zhou}}, \bibinfo {author}
  {\bibfnamefont {D.}~\bibnamefont {Wang}}, \bibinfo {author} {\bibfnamefont
  {X.}~\bibnamefont {Liu}}, \bibinfo {author} {\bibfnamefont {W.}~\bibnamefont
  {Zhao}},\ and\ \bibinfo {author} {\bibfnamefont {M.}~\bibnamefont {Ezawa}},\
  }\bibfield  {title} {\bibinfo {title} {{Control and manipulation of a
  magnetic skyrmionium in nanostructures}},\ }\bibfield  {journal} {\bibinfo
  {journal} {Physical Review B}\ }\textbf {\bibinfo {volume} {94}},\ \href
  {https://doi.org/10.1103/PhysRevB.94.094420} {10.1103/PhysRevB.94.094420}
  (\bibinfo {year} {2016}{\natexlab{a}})\BibitemShut {NoStop}%
\bibitem [{\citenamefont {Tomasello}\ \emph {et~al.}(2015)\citenamefont
  {Tomasello}, \citenamefont {Martinez}, \citenamefont {Zivieri}, \citenamefont
  {Torres}, \citenamefont {Carpentieri},\ and\ \citenamefont {Finocchio}}]{5}%
  \BibitemOpen
  \bibfield  {author} {\bibinfo {author} {\bibfnamefont {R.}~\bibnamefont
  {Tomasello}}, \bibinfo {author} {\bibfnamefont {E.}~\bibnamefont {Martinez}},
  \bibinfo {author} {\bibfnamefont {R.}~\bibnamefont {Zivieri}}, \bibinfo
  {author} {\bibfnamefont {L.}~\bibnamefont {Torres}}, \bibinfo {author}
  {\bibfnamefont {M.}~\bibnamefont {Carpentieri}},\ and\ \bibinfo {author}
  {\bibfnamefont {G.}~\bibnamefont {Finocchio}},\ }\bibfield  {title} {\bibinfo
  {title} {{A strategy for the design of skyrmion racetrack memories}},\ }\href
  {https://doi.org/10.1038/srep06784} {\bibfield  {journal} {\bibinfo
  {journal} {Scientific Reports}\ }\textbf {\bibinfo {volume} {4}},\ \bibinfo
  {pages} {6784} (\bibinfo {year} {2015})}\BibitemShut {NoStop}%
\bibitem [{\citenamefont {Yu}\ \emph {et~al.}(2016)\citenamefont {Yu},
  \citenamefont {Upadhyaya}, \citenamefont {Li}, \citenamefont {Li},
  \citenamefont {Kim}, \citenamefont {Fan}, \citenamefont {Wong}, \citenamefont
  {Tserkovnyak}, \citenamefont {Amiri},\ and\ \citenamefont
  {Wang}}]{Yu2016Room-TemperatureAsymmetry}%
  \BibitemOpen
  \bibfield  {author} {\bibinfo {author} {\bibfnamefont {G.}~\bibnamefont
  {Yu}}, \bibinfo {author} {\bibfnamefont {P.}~\bibnamefont {Upadhyaya}},
  \bibinfo {author} {\bibfnamefont {X.}~\bibnamefont {Li}}, \bibinfo {author}
  {\bibfnamefont {W.}~\bibnamefont {Li}}, \bibinfo {author} {\bibfnamefont
  {S.~K.}\ \bibnamefont {Kim}}, \bibinfo {author} {\bibfnamefont
  {Y.}~\bibnamefont {Fan}}, \bibinfo {author} {\bibfnamefont {K.~L.}\
  \bibnamefont {Wong}}, \bibinfo {author} {\bibfnamefont {Y.}~\bibnamefont
  {Tserkovnyak}}, \bibinfo {author} {\bibfnamefont {P.~K.}\ \bibnamefont
  {Amiri}},\ and\ \bibinfo {author} {\bibfnamefont {K.~L.}\ \bibnamefont
  {Wang}},\ }\bibfield  {title} {\bibinfo {title} {{Room-Temperature Creation
  and Spin-Orbit Torque Manipulation of Skyrmions in Thin Films with Engineered
  Asymmetry}},\ }\href {https://doi.org/10.1021/acs.nanolett.5b05257}
  {\bibfield  {journal} {\bibinfo  {journal} {Nano Letters}\ }\textbf {\bibinfo
  {volume} {16}},\ \bibinfo {pages} {1981} (\bibinfo {year}
  {2016})}\BibitemShut {NoStop}%
\bibitem [{\citenamefont {Tejo}\ \emph {et~al.}(2018)\citenamefont {Tejo},
  \citenamefont {Riveros}, \citenamefont {Escrig}, \citenamefont {Guslienko},\
  and\ \citenamefont {Chubykalo-Fesenko}}]{Tejo2017}%
  \BibitemOpen
  \bibfield  {author} {\bibinfo {author} {\bibfnamefont {F.}~\bibnamefont
  {Tejo}}, \bibinfo {author} {\bibfnamefont {A.}~\bibnamefont {Riveros}},
  \bibinfo {author} {\bibfnamefont {J.}~\bibnamefont {Escrig}}, \bibinfo
  {author} {\bibfnamefont {K.~Y.}\ \bibnamefont {Guslienko}},\ and\ \bibinfo
  {author} {\bibfnamefont {O.}~\bibnamefont {Chubykalo-Fesenko}},\ }\bibfield
  {title} {\bibinfo {title} {{Distinct magnetic field dependence of N{\'{e}}el
  skyrmion sizes in ultrathin nanodots}},\ }\href
  {https://doi.org/10.1038/s41598-018-24582-x} {\bibfield  {journal} {\bibinfo
  {journal} {Scientific Reports}\ }\textbf {\bibinfo {volume} {8}},\ \bibinfo
  {pages} {6280} (\bibinfo {year} {2018})}\BibitemShut {NoStop}%
\bibitem [{\citenamefont {Saha}\ \emph {et~al.}(2019)\citenamefont {Saha},
  \citenamefont {Zelent}, \citenamefont {Finizio}, \citenamefont
  {Mruczkiewicz}, \citenamefont {Tacchi}, \citenamefont {Suszka}, \citenamefont
  {Wintz}, \citenamefont {Bingham}, \citenamefont {Raabe}, \citenamefont
  {Krawczyk},\ and\ \citenamefont {Heyderman}}]{Saha2019FormationAnisotropy}%
  \BibitemOpen
  \bibfield  {author} {\bibinfo {author} {\bibfnamefont {S.}~\bibnamefont
  {Saha}}, \bibinfo {author} {\bibfnamefont {M.}~\bibnamefont {Zelent}},
  \bibinfo {author} {\bibfnamefont {S.}~\bibnamefont {Finizio}}, \bibinfo
  {author} {\bibfnamefont {M.}~\bibnamefont {Mruczkiewicz}}, \bibinfo {author}
  {\bibfnamefont {S.}~\bibnamefont {Tacchi}}, \bibinfo {author} {\bibfnamefont
  {A.~K.}\ \bibnamefont {Suszka}}, \bibinfo {author} {\bibfnamefont
  {S.}~\bibnamefont {Wintz}}, \bibinfo {author} {\bibfnamefont {N.~S.}\
  \bibnamefont {Bingham}}, \bibinfo {author} {\bibfnamefont {J.}~\bibnamefont
  {Raabe}}, \bibinfo {author} {\bibfnamefont {M.}~\bibnamefont {Krawczyk}},\
  and\ \bibinfo {author} {\bibfnamefont {L.~J.}\ \bibnamefont {Heyderman}},\
  }\bibfield  {title} {\bibinfo {title} {{Formation of N{\'{e}}el-type
  skyrmions in an antidot lattice with perpendicular magnetic anisotropy}},\
  }\href {https://doi.org/10.1103/PhysRevB.100.144435} {\bibfield  {journal}
  {\bibinfo  {journal} {Physical Review B}\ }\textbf {\bibinfo {volume}
  {100}},\ \bibinfo {pages} {144435} (\bibinfo {year} {2019})}\BibitemShut
  {NoStop}%
\bibitem [{\citenamefont {Zelent}\ \emph {et~al.}(2017)\citenamefont {Zelent},
  \citenamefont {T{\'{o}}bik}, \citenamefont {Krawczyk}, \citenamefont
  {Guslienko},\ and\ \citenamefont {Mruczkiewicz}}]{PSSR:PSSR201700259}%
  \BibitemOpen
  \bibfield  {author} {\bibinfo {author} {\bibfnamefont {M.}~\bibnamefont
  {Zelent}}, \bibinfo {author} {\bibfnamefont {J.}~\bibnamefont {T{\'{o}}bik}},
  \bibinfo {author} {\bibfnamefont {M.}~\bibnamefont {Krawczyk}}, \bibinfo
  {author} {\bibfnamefont {K.~Y.}\ \bibnamefont {Guslienko}},\ and\ \bibinfo
  {author} {\bibfnamefont {M.}~\bibnamefont {Mruczkiewicz}},\ }\bibfield
  {title} {\bibinfo {title} {{Bi-Stability of Magnetic Skyrmions in Ultrathin
  Multilayer Nanodots Induced by Magnetostatic Interaction}},\ }\href
  {https://doi.org/10.1002/pssr.201700259} {\bibfield  {journal} {\bibinfo
  {journal} {physica status solidi (RRL) - Rapid Research Letters}\ }\textbf
  {\bibinfo {volume} {11}},\ \bibinfo {pages} {1700259} (\bibinfo {year}
  {2017})}\BibitemShut {NoStop}%
\bibitem [{\citenamefont {Vetrova}\ \emph {et~al.}(2021)\citenamefont
  {Vetrova}, \citenamefont {Zelent}, \citenamefont {{\v{S}}olt{\'{y}}s},
  \citenamefont {Gubanov}, \citenamefont {Sadovnikov}, \citenamefont
  {{\v{S}}cepka}, \citenamefont {D{\'{e}}rer}, \citenamefont {Stoklas},
  \citenamefont {Cambel},\ and\ \citenamefont
  {Mruczkiewicz}}]{Vetrova2021InvestigationDot}%
  \BibitemOpen
  \bibfield  {author} {\bibinfo {author} {\bibfnamefont {I.~V.}\ \bibnamefont
  {Vetrova}}, \bibinfo {author} {\bibfnamefont {M.}~\bibnamefont {Zelent}},
  \bibinfo {author} {\bibfnamefont {J.}~\bibnamefont {{\v{S}}olt{\'{y}}s}},
  \bibinfo {author} {\bibfnamefont {V.~A.}\ \bibnamefont {Gubanov}}, \bibinfo
  {author} {\bibfnamefont {A.~V.}\ \bibnamefont {Sadovnikov}}, \bibinfo
  {author} {\bibfnamefont {T.}~\bibnamefont {{\v{S}}cepka}}, \bibinfo {author}
  {\bibfnamefont {J.}~\bibnamefont {D{\'{e}}rer}}, \bibinfo {author}
  {\bibfnamefont {R.}~\bibnamefont {Stoklas}}, \bibinfo {author} {\bibfnamefont
  {V.}~\bibnamefont {Cambel}},\ and\ \bibinfo {author} {\bibfnamefont
  {M.}~\bibnamefont {Mruczkiewicz}},\ }\bibfield  {title} {\bibinfo {title}
  {{Investigation of self-nucleated skyrmion states in the
  ferromagnetic/nonmagnetic multilayer dot}},\ }\href
  {https://doi.org/10.1063/5.0045835} {\bibfield  {journal} {\bibinfo
  {journal} {Applied Physics Letters}\ }\textbf {\bibinfo {volume} {118}},\
  \bibinfo {pages} {212409} (\bibinfo {year} {2021})}\BibitemShut {NoStop}%
\bibitem [{\citenamefont {Riveros}\ \emph {et~al.}(2021)\citenamefont
  {Riveros}, \citenamefont {Tejo}, \citenamefont {Escrig}, \citenamefont
  {Guslienko},\ and\ \citenamefont
  {Chubykalo-Fesenko}}]{Riveros2021Field-DependentNanodots}%
  \BibitemOpen
  \bibfield  {author} {\bibinfo {author} {\bibfnamefont {A.}~\bibnamefont
  {Riveros}}, \bibinfo {author} {\bibfnamefont {F.}~\bibnamefont {Tejo}},
  \bibinfo {author} {\bibfnamefont {J.}~\bibnamefont {Escrig}}, \bibinfo
  {author} {\bibfnamefont {K.~Y.}\ \bibnamefont {Guslienko}},\ and\ \bibinfo
  {author} {\bibfnamefont {O.}~\bibnamefont {Chubykalo-Fesenko}},\ }\bibfield
  {title} {\bibinfo {title} {{Field-Dependent Energy Barriers of Magnetic Neel
  Skyrmions in Ultrathin Circular Nanodots}},\ }\href
  {https://doi.org/10.1103/PHYSREVAPPLIED.16.014068/FIGURES/10/MEDIUM}
  {\bibfield  {journal} {\bibinfo  {journal} {Physical Review Applied}\
  }\textbf {\bibinfo {volume} {16}},\ \bibinfo {pages} {014068} (\bibinfo
  {year} {2021})}\BibitemShut {NoStop}%
\bibitem [{\citenamefont {Rajib}\ \emph {et~al.}(2021)\citenamefont {Rajib},
  \citenamefont {Misba}, \citenamefont {Bhattacharya},\ and\ \citenamefont
  {Atulasimha}}]{Rajib2021RobustDMI}%
  \BibitemOpen
  \bibfield  {author} {\bibinfo {author} {\bibfnamefont {M.~M.}\ \bibnamefont
  {Rajib}}, \bibinfo {author} {\bibfnamefont {W.~A.}\ \bibnamefont {Misba}},
  \bibinfo {author} {\bibfnamefont {D.}~\bibnamefont {Bhattacharya}},\ and\
  \bibinfo {author} {\bibfnamefont {J.}~\bibnamefont {Atulasimha}},\ }\bibfield
   {title} {\bibinfo {title} {{Robust skyrmion mediated reversal of
  ferromagnetic nanodots of 20 nm lateral dimension with high Ms and
  observable DMI}},\ }\href {https://doi.org/10.1038/s41598-021-99780-1}
  {\bibfield  {journal} {\bibinfo  {journal} {Scientific Reports}\ }\textbf
  {\bibinfo {volume} {11}},\ \bibinfo {pages} {20914} (\bibinfo {year}
  {2021})}\BibitemShut {NoStop}%
\bibitem [{\citenamefont {Zhang}\ \emph
  {et~al.}(2016{\natexlab{b}})\citenamefont {Zhang}, \citenamefont {Zhou},\
  and\ \citenamefont {Ezawa}}]{Zhang2016}%
  \BibitemOpen
  \bibfield  {author} {\bibinfo {author} {\bibfnamefont {X.}~\bibnamefont
  {Zhang}}, \bibinfo {author} {\bibfnamefont {Y.}~\bibnamefont {Zhou}},\ and\
  \bibinfo {author} {\bibfnamefont {M.}~\bibnamefont {Ezawa}},\ }\bibfield
  {title} {\bibinfo {title} {{Magnetic bilayer-skyrmions without skyrmion Hall
  effect}},\ }\href {https://doi.org/10.1038/ncomms10293
  https://www.nature.com/articles/ncomms10293{\#}supplementary-information}
  {\bibfield  {journal} {\bibinfo  {journal} {Nature Communications}\ }\textbf
  {\bibinfo {volume} {7}},\ \bibinfo {pages} {10293} (\bibinfo {year}
  {2016}{\natexlab{b}})}\BibitemShut {NoStop}%
\bibitem [{\citenamefont {Hsu}\ \emph {et~al.}(2016)\citenamefont {Hsu},
  \citenamefont {Kubetzka}, \citenamefont {Finco}, \citenamefont {Romming},
  \citenamefont {Bergmann},\ and\ \citenamefont {Wiesendanger}}]{Hsu2016}%
  \BibitemOpen
  \bibfield  {author} {\bibinfo {author} {\bibfnamefont {P.~J.}\ \bibnamefont
  {Hsu}}, \bibinfo {author} {\bibfnamefont {A.}~\bibnamefont {Kubetzka}},
  \bibinfo {author} {\bibfnamefont {A.}~\bibnamefont {Finco}}, \bibinfo
  {author} {\bibfnamefont {N.}~\bibnamefont {Romming}}, \bibinfo {author}
  {\bibfnamefont {K.~V.}\ \bibnamefont {Bergmann}},\ and\ \bibinfo {author}
  {\bibfnamefont {R.}~\bibnamefont {Wiesendanger}},\ }\bibfield  {title}
  {\bibinfo {title} {Electric-field-driven switching of individual magnetic
  skyrmions},\ }\href {https://doi.org/10.1038/nnano.2016.234} {\bibfield
  {journal} {\bibinfo  {journal} {Nature Nanotechnology 2016 12:2}\ }\textbf
  {\bibinfo {volume} {12}},\ \bibinfo {pages} {123} (\bibinfo {year}
  {2016})}\BibitemShut {NoStop}%
\bibitem [{\citenamefont {Jena}\ \emph {et~al.}(2020)\citenamefont {Jena},
  \citenamefont {G{\"{o}}bel}, \citenamefont {Kumar}, \citenamefont {Mertig},
  \citenamefont {Felser},\ and\ \citenamefont
  {Parkin}}]{Jena2020EvolutionSymmetry}%
  \BibitemOpen
  \bibfield  {author} {\bibinfo {author} {\bibfnamefont {J.}~\bibnamefont
  {Jena}}, \bibinfo {author} {\bibfnamefont {B.}~\bibnamefont {G{\"{o}}bel}},
  \bibinfo {author} {\bibfnamefont {V.}~\bibnamefont {Kumar}}, \bibinfo
  {author} {\bibfnamefont {I.}~\bibnamefont {Mertig}}, \bibinfo {author}
  {\bibfnamefont {C.}~\bibnamefont {Felser}},\ and\ \bibinfo {author}
  {\bibfnamefont {S.}~\bibnamefont {Parkin}},\ }\bibfield  {title} {\bibinfo
  {title} {{Evolution and competition between chiral spin textures in
  nanostripes with D2d symmetry}},\ }\bibfield  {journal} {\bibinfo  {journal}
  {Science Advances}\ }\textbf {\bibinfo {volume} {6}},\ \href
  {https://doi.org/10.1126/sciadv.abc0723} {10.1126/sciadv.abc0723} (\bibinfo
  {year} {2020})\BibitemShut {NoStop}%
\bibitem [{\citenamefont {Xia}\ \emph {et~al.}(2020)\citenamefont {Xia},
  \citenamefont {Zhang}, \citenamefont {Ezawa}, \citenamefont {Shao},
  \citenamefont {Liu},\ and\ \citenamefont {Zhou}}]{Xia2020}%
  \BibitemOpen
  \bibfield  {author} {\bibinfo {author} {\bibfnamefont {J.}~\bibnamefont
  {Xia}}, \bibinfo {author} {\bibfnamefont {X.}~\bibnamefont {Zhang}}, \bibinfo
  {author} {\bibfnamefont {M.}~\bibnamefont {Ezawa}}, \bibinfo {author}
  {\bibfnamefont {Q.}~\bibnamefont {Shao}}, \bibinfo {author} {\bibfnamefont
  {X.}~\bibnamefont {Liu}},\ and\ \bibinfo {author} {\bibfnamefont
  {Y.}~\bibnamefont {Zhou}},\ }\bibfield  {title} {\bibinfo {title} {Dynamics
  of an elliptical ferromagnetic skyrmion driven by the spinâ€“orbit
  torque},\ }\href {https://doi.org/10.1063/1.5132915} {\bibfield  {journal}
  {\bibinfo  {journal} {Applied Physics Letters}\ }\textbf {\bibinfo {volume}
  {116}},\ \bibinfo {pages} {022407} (\bibinfo {year} {2020})}\BibitemShut
  {NoStop}%
\bibitem [{\citenamefont {Zhang}\ \emph
  {et~al.}(2020{\natexlab{b}})\citenamefont {Zhang}, \citenamefont {Zhang},
  \citenamefont {Wen}, \citenamefont {Peng}, \citenamefont {Qiu}, \citenamefont
  {Matsumoto},\ and\ \citenamefont {Zhang}}]{Zhang2020DeformationMicroscopy}%
  \BibitemOpen
  \bibfield  {author} {\bibinfo {author} {\bibfnamefont {S.}~\bibnamefont
  {Zhang}}, \bibinfo {author} {\bibfnamefont {J.}~\bibnamefont {Zhang}},
  \bibinfo {author} {\bibfnamefont {Y.}~\bibnamefont {Wen}}, \bibinfo {author}
  {\bibfnamefont {Y.}~\bibnamefont {Peng}}, \bibinfo {author} {\bibfnamefont
  {Z.}~\bibnamefont {Qiu}}, \bibinfo {author} {\bibfnamefont {T.}~\bibnamefont
  {Matsumoto}},\ and\ \bibinfo {author} {\bibfnamefont {X.}~\bibnamefont
  {Zhang}},\ }\bibfield  {title} {\bibinfo {title} {{Deformation of
  N{\'{e}}el-type skyrmions revealed by Lorentz transmission electron
  microscopy}},\ }\href {https://doi.org/10.1063/5.0002592} {\bibfield
  {journal} {\bibinfo  {journal} {Applied Physics Letters}\ }\textbf {\bibinfo
  {volume} {116}},\ \bibinfo {pages} {142402} (\bibinfo {year}
  {2020}{\natexlab{b}})}\BibitemShut {NoStop}%
\bibitem [{\citenamefont {Shibata}\ \emph {et~al.}(2015)\citenamefont
  {Shibata}, \citenamefont {Iwasaki}, \citenamefont {Kanazawa}, \citenamefont
  {Aizawa}, \citenamefont {Tanigaki}, \citenamefont {Shirai}, \citenamefont
  {Nakajima}, \citenamefont {Kubota}, \citenamefont {Kawasaki}, \citenamefont
  {Park}, \citenamefont {Shindo}, \citenamefont {Nagaosa},\ and\ \citenamefont
  {Tokura}}]{Shibata2015LargeCrystal}%
  \BibitemOpen
  \bibfield  {author} {\bibinfo {author} {\bibfnamefont {K.}~\bibnamefont
  {Shibata}}, \bibinfo {author} {\bibfnamefont {J.}~\bibnamefont {Iwasaki}},
  \bibinfo {author} {\bibfnamefont {N.}~\bibnamefont {Kanazawa}}, \bibinfo
  {author} {\bibfnamefont {S.}~\bibnamefont {Aizawa}}, \bibinfo {author}
  {\bibfnamefont {T.}~\bibnamefont {Tanigaki}}, \bibinfo {author}
  {\bibfnamefont {M.}~\bibnamefont {Shirai}}, \bibinfo {author} {\bibfnamefont
  {T.}~\bibnamefont {Nakajima}}, \bibinfo {author} {\bibfnamefont
  {M.}~\bibnamefont {Kubota}}, \bibinfo {author} {\bibfnamefont
  {M.}~\bibnamefont {Kawasaki}}, \bibinfo {author} {\bibfnamefont {H.~S.}\
  \bibnamefont {Park}}, \bibinfo {author} {\bibfnamefont {D.}~\bibnamefont
  {Shindo}}, \bibinfo {author} {\bibfnamefont {N.}~\bibnamefont {Nagaosa}},\
  and\ \bibinfo {author} {\bibfnamefont {Y.}~\bibnamefont {Tokura}},\
  }\bibfield  {title} {\bibinfo {title} {{Large anisotropic deformation of
  skyrmions in strained crystal}},\ }\href
  {https://doi.org/10.1038/nnano.2015.113} {\bibfield  {journal} {\bibinfo
  {journal} {Nature Nanotechnology}\ }\textbf {\bibinfo {volume} {10}},\
  \bibinfo {pages} {589} (\bibinfo {year} {2015})}\BibitemShut {NoStop}%
\bibitem [{\citenamefont {Hagemeister}\ \emph {et~al.}(2016)\citenamefont
  {Hagemeister}, \citenamefont {Vedmedenko},\ and\ \citenamefont
  {Wiesendanger}}]{Hagemeister2016}%
  \BibitemOpen
  \bibfield  {author} {\bibinfo {author} {\bibfnamefont {J.}~\bibnamefont
  {Hagemeister}}, \bibinfo {author} {\bibfnamefont {E.~Y.}\ \bibnamefont
  {Vedmedenko}},\ and\ \bibinfo {author} {\bibfnamefont {R.}~\bibnamefont
  {Wiesendanger}},\ }\bibfield  {title} {\bibinfo {title} {Pattern formation in
  skyrmionic materials with anisotropic environments},\ }\href
  {https://doi.org/10.1103/PhysRevB.94.104434} {\bibfield  {journal} {\bibinfo
  {journal} {Phys. Rev. B}\ }\textbf {\bibinfo {volume} {94}},\ \bibinfo
  {pages} {104434} (\bibinfo {year} {2016})}\BibitemShut {NoStop}%
\bibitem [{\citenamefont {Cui}\ \emph {et~al.}(2021)\citenamefont {Cui},
  \citenamefont {Yu}, \citenamefont {Shao}, \citenamefont {Liu}, \citenamefont
  {Wu}, \citenamefont {Nan}, \citenamefont {Zhu}, \citenamefont {Wu},
  \citenamefont {Guo}, \citenamefont {Chen}, \citenamefont {Zhou},
  \citenamefont {Xi}, \citenamefont {Jiang}, \citenamefont {Wang},
  \citenamefont {Liang}, \citenamefont {Du}, \citenamefont {Wang},
  \citenamefont {Wang}, \citenamefont {Wu}, \citenamefont {Han}, \citenamefont
  {Zhang}, \citenamefont {Yang},\ and\ \citenamefont {Yu}}]{Cui2021}%
  \BibitemOpen
  \bibfield  {author} {\bibinfo {author} {\bibfnamefont {B.}~\bibnamefont
  {Cui}}, \bibinfo {author} {\bibfnamefont {D.}~\bibnamefont {Yu}}, \bibinfo
  {author} {\bibfnamefont {Z.}~\bibnamefont {Shao}}, \bibinfo {author}
  {\bibfnamefont {Y.}~\bibnamefont {Liu}}, \bibinfo {author} {\bibfnamefont
  {H.}~\bibnamefont {Wu}}, \bibinfo {author} {\bibfnamefont {P.}~\bibnamefont
  {Nan}}, \bibinfo {author} {\bibfnamefont {Z.}~\bibnamefont {Zhu}}, \bibinfo
  {author} {\bibfnamefont {C.}~\bibnamefont {Wu}}, \bibinfo {author}
  {\bibfnamefont {T.}~\bibnamefont {Guo}}, \bibinfo {author} {\bibfnamefont
  {P.}~\bibnamefont {Chen}}, \bibinfo {author} {\bibfnamefont {H.-A.}\
  \bibnamefont {Zhou}}, \bibinfo {author} {\bibfnamefont {L.}~\bibnamefont
  {Xi}}, \bibinfo {author} {\bibfnamefont {W.}~\bibnamefont {Jiang}}, \bibinfo
  {author} {\bibfnamefont {H.}~\bibnamefont {Wang}}, \bibinfo {author}
  {\bibfnamefont {S.}~\bibnamefont {Liang}}, \bibinfo {author} {\bibfnamefont
  {H.}~\bibnamefont {Du}}, \bibinfo {author} {\bibfnamefont {K.~L.}\
  \bibnamefont {Wang}}, \bibinfo {author} {\bibfnamefont {W.}~\bibnamefont
  {Wang}}, \bibinfo {author} {\bibfnamefont {K.}~\bibnamefont {Wu}}, \bibinfo
  {author} {\bibfnamefont {X.}~\bibnamefont {Han}}, \bibinfo {author}
  {\bibfnamefont {G.}~\bibnamefont {Zhang}}, \bibinfo {author} {\bibfnamefont
  {H.}~\bibnamefont {Yang}},\ and\ \bibinfo {author} {\bibfnamefont
  {G.}~\bibnamefont {Yu}},\ }\bibfield  {title} {\bibinfo {title}
  {N{\'e}el-type elliptical skyrmions in a laterally asymmetric magnetic
  multilayer},\ }\href {https://doi.org/https://doi.org/10.1002/adma.202006924}
  {\bibfield  {journal} {\bibinfo  {journal} {Advanced Materials}\ }\textbf
  {\bibinfo {volume} {33}},\ \bibinfo {pages} {2006924} (\bibinfo {year}
  {2021})}\BibitemShut {NoStop}%
\bibitem [{\citenamefont {Camosi}\ \emph {et~al.}(2021)\citenamefont {Camosi},
  \citenamefont {Garcia}, \citenamefont {Fruchart}, \citenamefont {Pizzini},
  \citenamefont {Locatelli}, \citenamefont {MenteÅŸ}, \citenamefont
  {Genuzio}, \citenamefont {Shaw}, \citenamefont {Nembach},\ and\ \citenamefont
  {Vogel}}]{Camosi2021}%
  \BibitemOpen
  \bibfield  {author} {\bibinfo {author} {\bibfnamefont {L.}~\bibnamefont
  {Camosi}}, \bibinfo {author} {\bibfnamefont {J.~P.}\ \bibnamefont {Garcia}},
  \bibinfo {author} {\bibfnamefont {O.}~\bibnamefont {Fruchart}}, \bibinfo
  {author} {\bibfnamefont {S.}~\bibnamefont {Pizzini}}, \bibinfo {author}
  {\bibfnamefont {A.}~\bibnamefont {Locatelli}}, \bibinfo {author}
  {\bibfnamefont {T.~O.}\ \bibnamefont {MenteÅŸ}}, \bibinfo {author}
  {\bibfnamefont {F.}~\bibnamefont {Genuzio}}, \bibinfo {author} {\bibfnamefont
  {J.~M.}\ \bibnamefont {Shaw}}, \bibinfo {author} {\bibfnamefont {H.~T.}\
  \bibnamefont {Nembach}},\ and\ \bibinfo {author} {\bibfnamefont
  {J.}~\bibnamefont {Vogel}},\ }\bibfield  {title} {\bibinfo {title}
  {Self-organised stripe domains and elliptical skyrmion bubbles in ultra-thin
  epitaxial au0.67pt0.33/co/w(110) films},\ }\href
  {https://doi.org/10.1088/1367-2630/ABDBE0} {\bibfield  {journal} {\bibinfo
  {journal} {New Journal of Physics}\ }\textbf {\bibinfo {volume} {23}},\
  \bibinfo {pages} {013020} (\bibinfo {year} {2021})}\BibitemShut {NoStop}%
\bibitem [{\citenamefont {Aranda}\ \emph {et~al.}(2018)\citenamefont {Aranda},
  \citenamefont {Hierro-Rodriguez}, \citenamefont {Kakazei}, \citenamefont
  {Chubykalo-Fesenko},\ and\ \citenamefont
  {Guslienko}}]{Aranda2018MagneticInteraction}%
  \BibitemOpen
  \bibfield  {author} {\bibinfo {author} {\bibfnamefont {A.~R.}\ \bibnamefont
  {Aranda}}, \bibinfo {author} {\bibfnamefont {A.}~\bibnamefont
  {Hierro-Rodriguez}}, \bibinfo {author} {\bibfnamefont {G.~N.}\ \bibnamefont
  {Kakazei}}, \bibinfo {author} {\bibfnamefont {O.}~\bibnamefont
  {Chubykalo-Fesenko}},\ and\ \bibinfo {author} {\bibfnamefont {K.~Y.}\
  \bibnamefont {Guslienko}},\ }\bibfield  {title} {\bibinfo {title} {{Magnetic
  skyrmion size and stability in ultrathin nanodots accounting
  Dzyaloshinskii-Moriya exchange interaction}},\ }\href
  {https://doi.org/10.1016/j.jmmm.2018.05.074} {\bibfield  {journal} {\bibinfo
  {journal} {Journal of Magnetism and Magnetic Materials}\ }\textbf {\bibinfo
  {volume} {465}},\ \bibinfo {pages} {471} (\bibinfo {year}
  {2018})}\BibitemShut {NoStop}%
\bibitem [{\citenamefont {Srivastava}\ \emph {et~al.}(2020)\citenamefont
  {Srivastava}, \citenamefont {Devi}, \citenamefont {Sharma}, \citenamefont
  {Ma}, \citenamefont {Deniz}, \citenamefont {Meyerheim}, \citenamefont
  {Felser}, \citenamefont {P~Parkin}, \citenamefont {Srivastava}, \citenamefont
  {Sharma}, \citenamefont {Ma}, \citenamefont {Deniz}, \citenamefont
  {Meyerheim}, \citenamefont {P~Parkin}, \citenamefont {Devi},\ and\
  \citenamefont {Felser}}]{Srivastava2020ObservationPtMnGa}%
  \BibitemOpen
  \bibfield  {author} {\bibinfo {author} {\bibfnamefont {A.~K.}\ \bibnamefont
  {Srivastava}}, \bibinfo {author} {\bibfnamefont {P.}~\bibnamefont {Devi}},
  \bibinfo {author} {\bibfnamefont {A.~K.}\ \bibnamefont {Sharma}}, \bibinfo
  {author} {\bibfnamefont {T.}~\bibnamefont {Ma}}, \bibinfo {author}
  {\bibfnamefont {H.}~\bibnamefont {Deniz}}, \bibinfo {author} {\bibfnamefont
  {H.~L.}\ \bibnamefont {Meyerheim}}, \bibinfo {author} {\bibfnamefont
  {C.}~\bibnamefont {Felser}}, \bibinfo {author} {\bibfnamefont {S.~S.}\
  \bibnamefont {P~Parkin}}, \bibinfo {author} {\bibfnamefont {A.~K.}\
  \bibnamefont {Srivastava}}, \bibinfo {author} {\bibfnamefont {A.~K.}\
  \bibnamefont {Sharma}}, \bibinfo {author} {\bibfnamefont {T.}~\bibnamefont
  {Ma}}, \bibinfo {author} {\bibfnamefont {H.}~\bibnamefont {Deniz}}, \bibinfo
  {author} {\bibfnamefont {H.~L.}\ \bibnamefont {Meyerheim}}, \bibinfo {author}
  {\bibfnamefont {S.~S.}\ \bibnamefont {P~Parkin}}, \bibinfo {author}
  {\bibfnamefont {P.}~\bibnamefont {Devi}},\ and\ \bibinfo {author}
  {\bibfnamefont {C.}~\bibnamefont {Felser}},\ }\bibfield  {title} {\bibinfo
  {title} {{Observation of Robust N{\'{e}}el Skyrmions in Metallic PtMnGa}},\
  }\href {https://doi.org/10.1002/ADMA.201904327} {\bibfield  {journal}
  {\bibinfo  {journal} {Advanced Materials}\ }\textbf {\bibinfo {volume}
  {32}},\ \bibinfo {pages} {1904327} (\bibinfo {year} {2020})}\BibitemShut
  {NoStop}%
\bibitem [{\citenamefont {Vansteenkiste}\ and\ \citenamefont
  {De~Wiele}(2011)}]{MuMax2011_main}%
  \BibitemOpen
  \bibfield  {author} {\bibinfo {author} {\bibfnamefont {A.}~\bibnamefont
  {Vansteenkiste}}\ and\ \bibinfo {author} {\bibfnamefont {B.~V.}\ \bibnamefont
  {De~Wiele}},\ }\bibfield  {title} {\bibinfo {title} {{MUMAX: A new
  high-performance micromagnetic simulation tool}},\ }\href
  {https://doi.org/10.1016/j.jmmm.2011.05.037} {\bibfield  {journal} {\bibinfo
  {journal} {Journal of Magnetism and Magnetic Materials}\ }\textbf {\bibinfo
  {volume} {323}},\ \bibinfo {pages} {2585} (\bibinfo {year}
  {2011})}\BibitemShut {NoStop}%
\bibitem [{\citenamefont {Vansteenkiste}\ \emph {et~al.}(2014)\citenamefont
  {Vansteenkiste}, \citenamefont {Leliaert}, \citenamefont {Dvornik},
  \citenamefont {Helsen}, \citenamefont {Garcia-Sanchez},\ and\ \citenamefont
  {Van~Waeyenberge}}]{mumax_2014}%
  \BibitemOpen
  \bibfield  {author} {\bibinfo {author} {\bibfnamefont {A.}~\bibnamefont
  {Vansteenkiste}}, \bibinfo {author} {\bibfnamefont {J.}~\bibnamefont
  {Leliaert}}, \bibinfo {author} {\bibfnamefont {M.}~\bibnamefont {Dvornik}},
  \bibinfo {author} {\bibfnamefont {M.}~\bibnamefont {Helsen}}, \bibinfo
  {author} {\bibfnamefont {F.}~\bibnamefont {Garcia-Sanchez}},\ and\ \bibinfo
  {author} {\bibfnamefont {B.}~\bibnamefont {Van~Waeyenberge}},\ }\bibfield
  {title} {\bibinfo {title} {{The design and verification of MuMax3}},\ }\href
  {https://doi.org/10.1063/1.4899186} {\bibfield  {journal} {\bibinfo
  {journal} {AIP Advances}\ }\textbf {\bibinfo {volume} {4}},\ \bibinfo {pages}
  {107133} (\bibinfo {year} {2014})}\BibitemShut {NoStop}%
\bibitem [{\citenamefont {Leliaert}\ \emph {et~al.}(2018)\citenamefont
  {Leliaert}, \citenamefont {Dvornik}, \citenamefont {Mulkers}, \citenamefont
  {De~Clercq}, \citenamefont {Milo{\v{s}}evi{\'{c}}},\ and\ \citenamefont
  {Van~Waeyenberge}}]{Leliaert2018FastMumax3}%
  \BibitemOpen
  \bibfield  {author} {\bibinfo {author} {\bibfnamefont {J.}~\bibnamefont
  {Leliaert}}, \bibinfo {author} {\bibfnamefont {M.}~\bibnamefont {Dvornik}},
  \bibinfo {author} {\bibfnamefont {J.}~\bibnamefont {Mulkers}}, \bibinfo
  {author} {\bibfnamefont {J.}~\bibnamefont {De~Clercq}}, \bibinfo {author}
  {\bibfnamefont {M.~V.}\ \bibnamefont {Milo{\v{s}}evi{\'{c}}}},\ and\ \bibinfo
  {author} {\bibfnamefont {B.}~\bibnamefont {Van~Waeyenberge}},\ }\bibfield
  {title} {\bibinfo {title} {{Fast micromagnetic simulations on GPU - Recent
  advances made with mumax3}},\ }\href
  {https://doi.org/10.1088/1361-6463/aaab1c} {\bibfield  {journal} {\bibinfo
  {journal} {Journal of Physics D: Applied Physics}\ }\textbf {\bibinfo
  {volume} {51}},\ \bibinfo {pages} {123002} (\bibinfo {year}
  {2018})}\BibitemShut {NoStop}%
\bibitem [{\citenamefont {Beg}\ \emph {et~al.}(2015)\citenamefont {Beg},
  \citenamefont {Carey}, \citenamefont {Wang}, \citenamefont
  {Cort{\'{e}}s-Ortu{\~{n}}o}, \citenamefont {Vousden}, \citenamefont
  {Bisotti}, \citenamefont {Albert}, \citenamefont {Chernyshenko},
  \citenamefont {Hovorka}, \citenamefont {Stamps},\ and\ \citenamefont
  {Fangohr}}]{Beg_2015_Ground_state_search_hysteretic_behaviour}%
  \BibitemOpen
  \bibfield  {author} {\bibinfo {author} {\bibfnamefont {M.}~\bibnamefont
  {Beg}}, \bibinfo {author} {\bibfnamefont {R.}~\bibnamefont {Carey}}, \bibinfo
  {author} {\bibfnamefont {W.}~\bibnamefont {Wang}}, \bibinfo {author}
  {\bibfnamefont {D.}~\bibnamefont {Cort{\'{e}}s-Ortu{\~{n}}o}}, \bibinfo
  {author} {\bibfnamefont {M.}~\bibnamefont {Vousden}}, \bibinfo {author}
  {\bibfnamefont {M.~A.}\ \bibnamefont {Bisotti}}, \bibinfo {author}
  {\bibfnamefont {M.}~\bibnamefont {Albert}}, \bibinfo {author} {\bibfnamefont
  {D.}~\bibnamefont {Chernyshenko}}, \bibinfo {author} {\bibfnamefont
  {O.}~\bibnamefont {Hovorka}}, \bibinfo {author} {\bibfnamefont {R.~L.}\
  \bibnamefont {Stamps}},\ and\ \bibinfo {author} {\bibfnamefont
  {H.}~\bibnamefont {Fangohr}},\ }\bibfield  {title} {\bibinfo {title} {{Ground
  state search, hysteretic behaviour, and reversal mechanism of skyrmionic
  textures in confined helimagnetic nanostructures}},\ }\href
  {https://doi.org/10.1038/srep17137} {\bibfield  {journal} {\bibinfo
  {journal} {Scientific Reports}\ }\textbf {\bibinfo {volume} {5}},\ \bibinfo
  {pages} {17137} (\bibinfo {year} {2015})}\BibitemShut {NoStop}%
\bibitem [{Note1()}]{Note1}%
  \BibitemOpen
  \bibinfo {note} {For this purpose, we have performed an analysis of the
  magnetostatic field distribution of the system under study with reduced size
  of the elementary cell, just to obtain precise spatial distributions of the
  field. We used here 0.75 $\times $0.75 $\times $0.15 nm cell
  size.}\BibitemShut {Stop}%
\bibitem [{\citenamefont {Aharoni}\ \emph {et~al.}(2000)\citenamefont
  {Aharoni}, \citenamefont {Pust},\ and\ \citenamefont {Kief}}]{Aharoni2000}%
  \BibitemOpen
  \bibfield  {author} {\bibinfo {author} {\bibfnamefont {A.}~\bibnamefont
  {Aharoni}}, \bibinfo {author} {\bibfnamefont {L.}~\bibnamefont {Pust}},\ and\
  \bibinfo {author} {\bibfnamefont {M.}~\bibnamefont {Kief}},\ }\bibfield
  {title} {\bibinfo {title} {{Comparing Theoretical Demagnetizing Factors with
  the Observed Saturation Process in Rectangular Shields}},\ }\href
  {https://doi.org/doi:http://dx.doi.org/10.1063/1.372771} {\bibfield
  {journal} {\bibinfo  {journal} {J. Appl. Phys.}\ }\textbf {\bibinfo {volume}
  {87}},\ \bibinfo {pages} {6564} (\bibinfo {year} {2000})}\BibitemShut
  {NoStop}%
\bibitem [{\citenamefont {Coey}(2010)}]{Coey2010}%
  \BibitemOpen
  \bibfield  {author} {\bibinfo {author} {\bibfnamefont {J.~M.~D.}\
  \bibnamefont {Coey}},\ }\href {https://doi.org/10.1017/CBO9780511845000}
  {\emph {\bibinfo {title} {Magnetism and Magnetic Materials}}}\ (\bibinfo
  {publisher} {Cambridge University Press},\ \bibinfo {year}
  {2010})\BibitemShut {NoStop}%
\bibitem [{\citenamefont {Chen}\ \emph {et~al.}(2021)\citenamefont {Chen},
  \citenamefont {Hu},\ and\ \citenamefont {Yu}}]{Chen2021ChiralSkyrmions}%
  \BibitemOpen
  \bibfield  {author} {\bibinfo {author} {\bibfnamefont {J.}~\bibnamefont
  {Chen}}, \bibinfo {author} {\bibfnamefont {J.}~\bibnamefont {Hu}},\ and\
  \bibinfo {author} {\bibfnamefont {H.}~\bibnamefont {Yu}},\ }\bibfield
  {title} {\bibinfo {title} {{Chiral Emission of Exchange Spin Waves by
  Magnetic Skyrmions}},\ }\href {https://doi.org/10.1021/acsnano.0c07805}
  {\bibfield  {journal} {\bibinfo  {journal} {ACS Nano}\ }\textbf {\bibinfo
  {volume} {15}},\ \bibinfo {pages} {4372} (\bibinfo {year}
  {2021})}\BibitemShut {NoStop}%
\bibitem [{\citenamefont {Marioni}\ \emph {et~al.}(2018)\citenamefont
  {Marioni}, \citenamefont {Penedo}, \citenamefont {Ba{\'{c}}ani},
  \citenamefont {Schwenk},\ and\ \citenamefont
  {Hug}}]{Marioni2018HalbachTextures}%
  \BibitemOpen
  \bibfield  {author} {\bibinfo {author} {\bibfnamefont {M.~A.}\ \bibnamefont
  {Marioni}}, \bibinfo {author} {\bibfnamefont {M.}~\bibnamefont {Penedo}},
  \bibinfo {author} {\bibfnamefont {M.}~\bibnamefont {Ba{\'{c}}ani}}, \bibinfo
  {author} {\bibfnamefont {J.}~\bibnamefont {Schwenk}},\ and\ \bibinfo {author}
  {\bibfnamefont {H.~J.}\ \bibnamefont {Hug}},\ }\bibfield  {title} {\bibinfo
  {title} {{Halbach Effect at the Nanoscale from Chiral Spin Textures}},\
  }\href {https://doi.org/10.1021/acs.nanolett.7b04802} {\bibfield  {journal}
  {\bibinfo  {journal} {Nano Letters}\ }\textbf {\bibinfo {volume} {18}},\
  \bibinfo {pages} {2263} (\bibinfo {year} {2018})}\BibitemShut {NoStop}%
\bibitem [{\citenamefont {Tetienne}\ \emph {et~al.}(2015)\citenamefont
  {Tetienne}, \citenamefont {Hingant}, \citenamefont {Mart{\'{i}}nez},
  \citenamefont {Rohart}, \citenamefont {Thiaville}, \citenamefont {Diez},
  \citenamefont {Garcia}, \citenamefont {Adam}, \citenamefont {Kim},
  \citenamefont {Roch}, \citenamefont {Miron}, \citenamefont {Gaudin},
  \citenamefont {Vila}, \citenamefont {Ocker}, \citenamefont {Ravelosona},\
  and\ \citenamefont {Jacques}}]{Tetienne2015TheNanomagnetometry}%
  \BibitemOpen
  \bibfield  {author} {\bibinfo {author} {\bibfnamefont {J.-P.}\ \bibnamefont
  {Tetienne}}, \bibinfo {author} {\bibfnamefont {T.}~\bibnamefont {Hingant}},
  \bibinfo {author} {\bibfnamefont {L.}~\bibnamefont {Mart{\'{i}}nez}},
  \bibinfo {author} {\bibfnamefont {S.}~\bibnamefont {Rohart}}, \bibinfo
  {author} {\bibfnamefont {A.}~\bibnamefont {Thiaville}}, \bibinfo {author}
  {\bibfnamefont {L.~H.}\ \bibnamefont {Diez}}, \bibinfo {author}
  {\bibfnamefont {K.}~\bibnamefont {Garcia}}, \bibinfo {author} {\bibfnamefont
  {J.-P.}\ \bibnamefont {Adam}}, \bibinfo {author} {\bibfnamefont {J.-V.}\
  \bibnamefont {Kim}}, \bibinfo {author} {\bibfnamefont {J.-F.}\ \bibnamefont
  {Roch}}, \bibinfo {author} {\bibfnamefont {I.}~\bibnamefont {Miron}},
  \bibinfo {author} {\bibfnamefont {G.}~\bibnamefont {Gaudin}}, \bibinfo
  {author} {\bibfnamefont {L.}~\bibnamefont {Vila}}, \bibinfo {author}
  {\bibfnamefont {B.}~\bibnamefont {Ocker}}, \bibinfo {author} {\bibfnamefont
  {D.}~\bibnamefont {Ravelosona}},\ and\ \bibinfo {author} {\bibfnamefont
  {V.}~\bibnamefont {Jacques}},\ }\bibfield  {title} {\bibinfo {title} {{The
  nature of domain walls in ultrathin ferromagnets revealed by scanning
  nanomagnetometry}},\ }\href {https://doi.org/10.1038/ncomms7733} {\bibfield
  {journal} {\bibinfo  {journal} {Nature Communications}\ }\textbf {\bibinfo
  {volume} {6}},\ \bibinfo {pages} {6733} (\bibinfo {year} {2015})}\BibitemShut
  {NoStop}%
\bibitem [{\citenamefont {Zhang}\ and\ \citenamefont {Li}(2004)}]{Zhang2004}%
  \BibitemOpen
  \bibfield  {author} {\bibinfo {author} {\bibfnamefont {S.}~\bibnamefont
  {Zhang}}\ and\ \bibinfo {author} {\bibfnamefont {Z.}~\bibnamefont {Li}},\
  }\bibfield  {title} {\bibinfo {title} {{Roles of nonequilibrium conduction
  electrons on the magnetization dynamics of ferromagnets}},\ }\href@noop {}
  {\bibfield  {journal} {\bibinfo  {journal} {Physical Review Letters}\
  }\textbf {\bibinfo {volume} {93}},\ \bibinfo {pages} {127204} (\bibinfo
  {year} {2004})}\BibitemShut {NoStop}%
\end{thebibliography}


%

\pagebreak
\appendix

\begin{center}
\textbf{\large Supplemental Materials:\\
\title{Stabilization and application of  N\'eel skyrmions in hybrid nanostructures}}
\end{center}

\setcounter{equation}{0}
\setcounter{figure}{0}
\setcounter{table}{0}
\setcounter{page}{1}
\makeatletter
\renewcommand{\theequation}{S\arabic{equation}}
\renewcommand{\thefigure}{S\arabic{figure}}
\renewcommand{\bibnumfmt}[1]{[S#1]}
\renewcommand{\citenumfont}[1]{S#1}

\begin{figure}[!htp]
 \includegraphics[width=\textwidth]{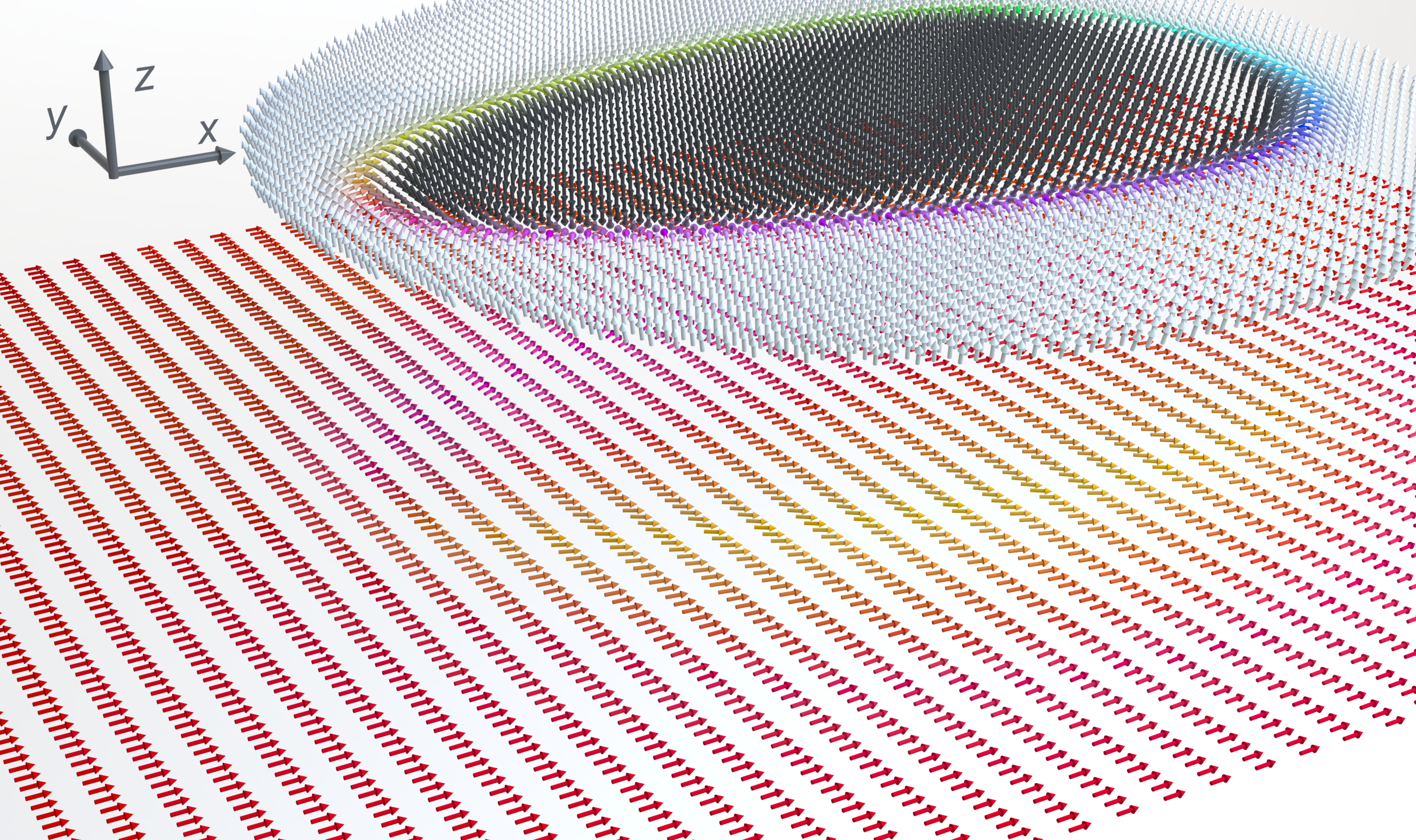}
 \caption{An artistic, high-resolution visualization of the simulated three-dimensional spatial distribution of the magnetic moments in nanodots and ferromagnetic stripes. 
 }
 \label{fig:SUP_sk_vis}
\end{figure}

To demonstrate the spatial distribution of the magnetic texture magnetization of N\'eel-type nanodot and the imprint in the stripe, a simulated high-resolution magnetic 3D spatial distribution of the mangetization texture is shown in Fig.~\ref{fig:SUP_sk_vis}. 

\begin{figure}
 \includegraphics[width=\textwidth]{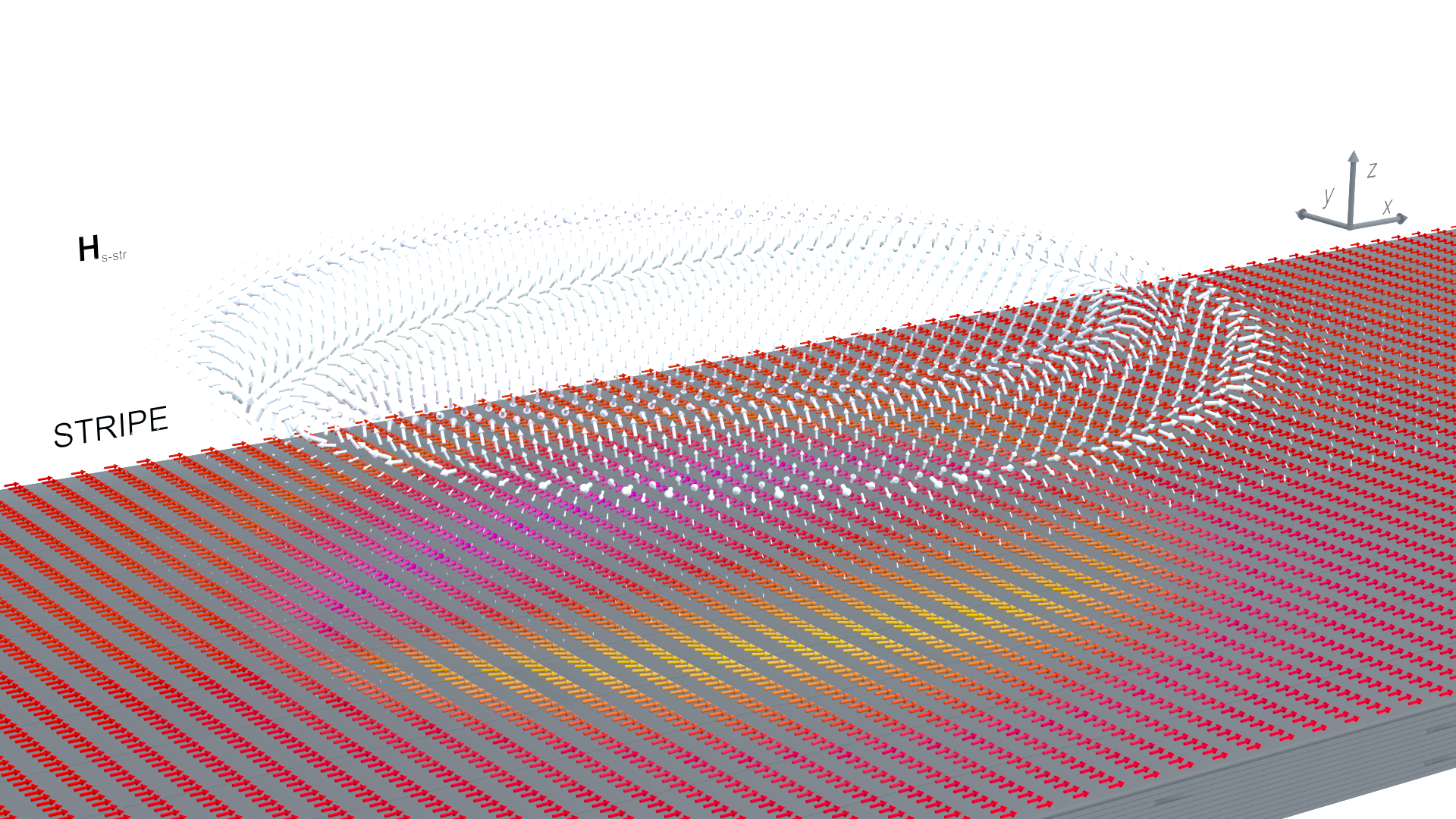}
 \caption{Magnetization in the Py stripe (coloured arrows) and distribution of the stray magnetic field (grey arrows) generated solely by the imprint in the stripe, excluding the field from the nanodot. The size of the arrow is proportional to the magnetic field strength. The simulation was performed with the freezing-spin technique with  DMI = -1.6 mJ/m$^{2}$.
 }
 \label{fig:SUP:b_demag_without_nd_zcross3}
\end{figure}

Figure ~\ref{fig:SUP:b_demag_without_nd_zcross3} presets a simulated high-resolution magnetic 3D spatial distribution of the magnetization texture and $\mathrm{\textbf{H}}_{\mathrm{s-str}}$ (x, y, z = 6.5 nm), the magnetostatic stray filed from the imprint. This stray field is opposite to and an order of magnitude lower than the field, which is created by the nanodot $\mathrm{\textbf{H}}_{\mathrm{s-dot}}$ .  

\begin{figure}[!htp]
\includegraphics[width=\textwidth]{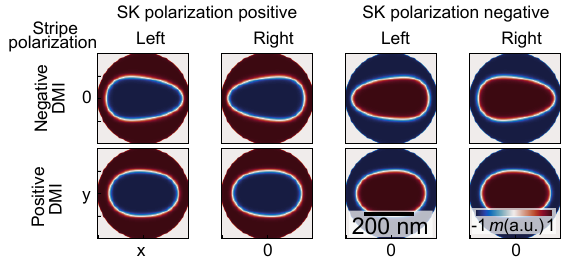}
 \caption{Static magnetization configuration in a Pt/Co/Ir nanodot with DMI = $\pm 1.6$ mJ/m$^2$ dipolarly coupled to the Py stripe. Static magnetization configuration in a Pt/Co/Ir nanodot with DMI = 1.6 mJ/m$^2$
dipolarly coupled to the Py stripe. The colour scale is given in reduced units of magnetization. The colour scale is given in reduced units of magnetization.
 }
 \label{fig:SUP_sk_polarization}
\end{figure}

To understand the reciprocal effects of skyrmion core polarization, DMI sign, and stripe magnetization direction on skyrmion shape and size, we performed a series of simulations for DMI = --1.6 mJ/m$^2$ of hybrid nanostructure (see Fig.~\ref{fig:SUP_sk_polarization}). Micromagnetic simulations showed that a horizontal change in the direction of magnetization orientation in the stripe changes the side of the narrower side of the skyrmion. Also an alteration in the skyrmion polarity in the stripe results in a horizontal mirror image of the skyrmion's magnetization texture in the nanodot.  No other effects were observed in the properties of skyrmion.   

\end{document}